\documentclass[aps,prd,11pt,a4paper,nofootinbib,oneside,superscriptaddress]{revtex4-1}
\pdfoutput = 1

\usepackage{amsmath,amssymb,amsfonts,color}
\usepackage{tensor,slashed,paralist,cases,mathrsfs}
\usepackage{float,cancel,xcolor}
\usepackage{graphicx}% Include figure files
\usepackage{dcolumn}% Align table columns on decimal point
\usepackage{bm}% bold math
\usepackage[colorlinks=true,urlcolor=blue,linkcolor=blue,citecolor=blue,linktocpage=true]{hyperref}

\begin{document}

\vspace*{1.5em}

\title{Search for Scalar Dark Matter via Pseudoscalar Portal Interactions: In Light of the Galactic Center Gamma-Ray Excess}

\author{Kwei-Chou Yang}
\email{kcyang@cycu.edu.tw}

\affiliation{Department of Physics and Center for High Energy Physics, Chung Yuan Christian University, Taoyuan 320, Taiwan}

%\date{\today $\vphantom{\bigg|_{\bigg|}^|}$}

\begin{abstract}

 In light of the observed Galactic center gamma-ray excess, we investigate a simplified model, for which the scalar dark matter interacts with quarks through a pseudoscalar mediator.  The viable regions of the parameter space, that can also account for the relic density and evade the current searches, are identified, if the low-velocity dark matter annihilates through an $s$-channel off shell mediator mostly into $\bar{b} b$, and/or annihilates directly into two {\it hidden} on shell mediators, which subsequently decay into the quark pairs. These two kinds of annihilations are $s$ wave.  
The projected monojet limit set by the high luminosity LHC sensitivity could constrain the favored parameter space, where the mediator's mass is larger than the dark matter mass by a factor of 2. We show that the projected sensitivity of 15-year Fermi-LAT observations of dwarf spheroidal galaxies can provide a stringent constraint on the most parameter space allowed in this model. If the on shell mediator channel contributes to the dark matter annihilation cross sections over 50\%,  this model with a lighter mediator can be probed in the projected PICO-500L experiment.

\end{abstract}
\maketitle
\newpage

\section{Introduction}

The standard model (SM) so far is  quite successfully tested in the current high energy physics experiments.  Nevertheless, the existence of the dark matter (DM) is indicated  by various astrophysical measurements and astronomical observations.  
The Galactic center (GC) is an excellent place to generate the DM signals, because it concentrates a large quantity of dark matter.  From the data collected by Fermi Large Area Telescope (Femi-LAT), several studies have found an excess of GeV gamma-rays near the region of the Galactic center \cite{Goodenough:2009gk, Hooper:2010mq, Hooper:2011ti, Abazajian:2012pn, Gordon:2013vta, Huang:2013pda, Daylan:2014rsa, Calore:2014xka, Calore:2014nla,Karwin:2016tsw,TheFermi-LAT:2017vmf}, where the spectrum and morphology can be interpreted as the signals generated by annihilating dark matter (DM) particles \cite{Goodenough:2009gk, Hooper:2010mq, Hooper:2011ti, Abazajian:2012pn, Gordon:2013vta, Huang:2013pda, Daylan:2014rsa,Calore:2014xka, Calore:2014nla,Karwin:2016tsw}. The interpretation is not conclusive yet. A population of millisecond pulsars (MSPs) has been proposed as a plausible origin of the GC gamma-ray excess \cite{Hooper:2015jlu,OLeary:2016cwz,Fermi-LAT:2017yoi,Ploeg:2017vai}. However, if so, the same region should contain a much more population of low-mass X-ray binaries than that observed so far \cite{Haggard:2017lyq}. On the other hand, Hooper {\it et al}. have also argued that if these MSPs convert more than a few percent of spin-down power into very high-energy $e^+ e^-$ pairs, then the inverse-Compton emission would exceed the observation by HESS \cite{Hooper:2017rzt}. 
 
 Although several interaction types of DM models could be responsible for the GC gamma-ray excess, the predictions for the DM mass and relevant parameters might be in tension with the stringent constraints from direct detection experiments and colliders.  A so-called ``coy dark matter" model recently stressed that, for the DM fermions interacting with SM particles via a pseudoscalar mediator~\cite{Boehm:2014hva,Buchmueller:2015eea,Hektor:2014kga,Izaguirre:2014vva,Ipek:2014gua, Arina:2014yna,Abdullah:2014lla,Cheung:2014lqa,Huang:2014cla,Cahill-Rowley:2014ora,Dolan:2014ska,Kozaczuk:2015bea,Kim:2016csm,Hektor:2016uth,Hektor:2017ftg,Banerjee:2017wxi, Cao:2014efa,Cao:2015loa,Baek:2017vzd,Berlin:2015wwa,No:2015xqa,Goncalves:2016iyg, Bauer:2017ota,Tunney:2017yfp}, the DM annihilation cross section into $b$ quarks can be large enough to provide a good fit to the GC gamma-ray excess, while its corresponding $t$-channel process (the DM-nucleus scattering), relevant to the direct detection, is suppressed by four powers of momentum transfer, and,
on the other hand, only a limited portion of the allowed parameter region can be constrained at the 14 TeV LHC run~\cite{Boehm:2014hva,Buchmueller:2015eea}.

Another idea, similar to the secluded dark matter scenario \cite{Profumo:2017obk}, is to introduce a model with hidden sector mediators, in which the DM first annihilates into on-shell mediators, and subsequently the mediators decay to the SM particles via a very small coupling~\cite{Abdullah:2014lla,Martin:2014sxa,Berlin:2014pya,Escudero:2017yia}. Because the low-velocity annihilation cross section is highly insensitive to the mediator's coupling to the SM particles, it can thus explain the GC gamma-ray excess and easily evade the stringent constraints from the direct detection and collider experiments.

Motivated by these results as mentioned above, we consider the scalar DM interactions with SM quarks via a pseudoscalar mediator. A pseudoscalar particle is interesting from a model-building point of view, because a  model with an extension of the Higgs sector, {\it e.g.} a two-Higgs doublet model, can naturally contain a such state.  For this model, interactions through the $s$-channel exchange of a pseudoscalar  with SM quarks could account for the GC gamma-ray excess \cite{Berlin:2014tja,Escudero:2016kpw}, and constraints from current direct detection experiments and the LHC could be obviated.
We find that, if the annihilation is only given by the pseudoscalar mediated $s$-channel process, the parameter region $m_A\lesssim2 m_\phi$, where $m_A$ and $m_\phi$ are the masses of the mediator and dark matter, respectively, is ruled out by a combination analysis of various experiments, especially by the relic density constraint and observations of dwarf spheroidal galaxies (dSphs)~\cite{Ackermann:2015zua}. The viable regions of the parameter space will be discussed.

On the other hand, because not only the scalar DM $s$-channel annihilation but also the DM annihilation into two {\it hidden} on-shell pseudoscalar mediators, via $t$ and $u$ channels, are $s$-wave processes, we find that, for $m_A<m_\phi$,  broad parameter regions that can provide a good fit to the observed GC gamma-ray excess and evade the current searches.
Unlike the previous works, where the hidden sector mediator model is characterized by very small values of the mediator-SM couplings, the strong suppression of signals for direct detections and colliders in the present hidden sector is mainly due to the coupling structure for which the mediator interacts with the SM quarks via pseudoscalar couplings.
We will show that, for  $m_A < m_\phi$, the DM annihilation into on-shell mediators gives sizable contributions to the observed GC gamma-ray excess.

In this paper, we adopt the framework of the simplified model, where the scalar dark matter annihilates through a spin-0 mediator with pseudoscalar couplings to SM quarks, which are assumed to be proportional to the Yukawa couplings motivated by minimal flavor violation~\cite{D'Ambrosio:2002ex}.   Using a minimal set of parameters, the simplified model can not only capture the feature of a specific ultraviolet (UV) complete model, but also provide a generic framework to perform the experimental data analysis.

We further consider the updated and projected bounds set by the gamma-ray observations of dSphs~\cite{Ackermann:2015zua}, direct detection experiments,  and LHC monojet result \cite{Sirunyan:2017hci}. The dSphs, containing little dust or gas, are believed to be DM dominated.  So far, no gamma-ray emission has been measured  from dSphs by Fermi-LAT.  For the direct detections, because the DM-nucleus interaction in this model is spin dependent, the LUX \cite{Akerib:2016vxi} signals mostly arise from the unpaired neutrons inside the abundant Xe isotopes, which are the LUX detector materials,  while the PICO-60~\cite{Amole:2015pla,Amole:2017dex,Amole:2016pye,Amole:2015lsj}, using the $\text{CF}_3\text{I}$ and $\text{C}_3\text{F}_8$ as targets, detects the signals mainly via unpaired protons inside the target nuclei. We find that in the direct searches the PICO results set the most stringent bound which is also insensitive to the choice of the parameter set.   At the LHC, the production of the scalar DM pair via the spin-0 mediator in the monojet accompanied by the missing transverse energy ($\not\!\!{E}_{\rm T}$) is dominated by the gluon fusion processes.  We give the monojet constraint on the relevant parameter space, by taking into account  the recent CMS 13 TeV results with 12.9 fb$^{-1}$ and projected sensitivity for the high luminosity LHC. 

The layout of this paper is as follows. In Sec.~\ref{sec:DM-model}, we introduce the simplified scalar DM model with a spin-0 mediator that couples to SM quarks via pseudoscalar couplings. This section includes the formulas about decay widths of the mediator, that are relevant to calculations for the DM relic density, indirect detection, and monojet result at the LHC. In Sec.~\ref{description_expt}, we describe the approaches in detail for model constraints due to observations, and their implementations.  For this model, the constraints on the parameter space are presented in Sec.~\ref{sec:constraints}. The conclusions are summarized in Sec.~\ref{sec:conclusions}.  The brief descriptions for the relic abundance and results of thermally averaged annihilation cross sections are given in Appendices \ref{app:relic} and \ref{app:annXS}, respectively.

\section{The Scalar Dark Matter Model}\label{sec:DM-model}

We consider a simplified model, where the scalar dark matter ($\phi^{(*)}$) annihilates through a spin-0 mediator ($A$) with pseudoscalar couplings to SM quarks ($q$). The effective Lagrangian is given by
\begin{eqnarray}\label{eq:ex-lagrangian-1}
{\cal L}_{\rm int}\supset  g_\phi m_\phi A \phi^*\phi +  i A \sum_q g_q \bar q  \gamma^5 q \,,
\end{eqnarray} 
for the complex scalar DM, or
\begin{eqnarray}\label{eq:ex-lagrangian-2}
{\cal L}_{\rm int}\supset  \frac{1}{2} g_\phi m_\phi A \phi^2 +  i A \sum_q g_q \bar q  \gamma^5 q \,,
\end{eqnarray} 
for the real scalar DM, where the latter one with the factor of $1/2$ gives the identical expression for the direct detection rate and annihilation cross section as the former case.
Motivated by the minimal flavor violation ansatz \cite{D'Ambrosio:2002ex}, we assume  the pseudoscalar-SM quark couplings are proportional to the associated Yukawa couplings, given by  $g_{q}=g \sqrt{2}m_q/v$ with $m_q$ being quark's mass and $v=246$ GeV. 
For simplicity, in the following, we consider the DM particle to be a real scalar, while the generalization to  a complex scalar one is straightforward.
Because this work is motivated by the fermionic case, we will assume the mediator is a pseudoscalar, such that the $A$-$\phi^{(*)}$-$\phi$ coupling is $CP$ violating; however, our conclusions are independent of the mediator's parity.  
For a pseudoscalar mediator, the UV completion of the simplified model  could be built up to relate to the Higgs-extension portal \cite{Ipek:2014gua, Cao:2014efa, Cao:2015loa,Berlin:2015wwa, No:2015xqa, Goncalves:2016iyg, Bauer:2017ota, Tunney:2017yfp, Bell:2017rgi} or axion-portal DM models \cite{Freytsis:2009ct}, which might contain more (mediator) particles in addition to an extra pseudoscalar compared with the SM and thus have richer phenomenologies (see also discussions in Refs.~\cite{Tunney:2017yfp, Bell:2017rgi} and LHC constraints in Refs.~\cite{Berlin:2015wwa,No:2015xqa, Goncalves:2016iyg, Bauer:2017ota}).

We have collected the results for the annihilation cross sections, that are relevant to the relic density and indirect detection searches, in Appendix \ref{app:annXS}. The  partial decay widths of the  pseudoscalar particle are
\begin{align}
\Gamma(A \rightarrow \bar{q} q) & = g_q^2 \frac{3 m_A}{8\pi}  \left( 1- \frac{4 m_q^2}{m_A^2} \right)^{1/2} \theta(m_A-2m_q) \; ,  \label{eq:partial-width-1} \\
\Gamma(A \rightarrow g g) & = \frac{\alpha_s^2}{2  \pi^3 m_A}\left|\sum_q m_q g_q f_A \left( \frac{4 m_q^2}{m_A^2}\right)   \right|^2 \; , \label{eq:partial-width-2}  \\
\Gamma(A \rightarrow \phi \phi) & = g_\phi^2 \frac{m_\phi^2}{32 \pi m_A}  \left( 1- \frac{4 m_\phi^2}{m_A^2} \right)^{1/2} \theta(m_A-2m_\phi) \label{eq:partial-width-3}   \; ,
\end{align}
where 
\begin{equation}
f_A(\tau)  =  \left\{ \begin{array}{lr} \text{arcsin}^2 \sqrt{\tau^{-1}}
        \, , \ \ & \tau \geq 1 \\ -\frac{1}{4} \left[ \log \frac{1+\sqrt{1-\tau}}{1-\sqrt{1-\tau}} - i \pi\right]^2 \, , & \tau < 1 \end{array} \right. \; .
\end{equation}
The DM particles annihilating into the quark pair, via the $s$ channel,  is $s$ wave, while if kinematically accessible ($m_\phi > m_A$), the annihilation, via $t$ and $u$ channels, into  on-shell mediators,  is also $s$ wave. We will show that the latter cannot be neglected, when it is allowed.
 Motivated by the phenomenological considerations, we choose the following three scenarios to explain the GC gamma-ray excess: (i) scenario {\bf s1}:  completely due to the DM annihilations into quark pairs, i.e., $\sum_q \langle \sigma v\rangle_{q\bar{q}}$,   (ii) scenario {\bf s2}:  due to the annihilations equally into quark pairs and into on-shell mediators, i.e.,  $\sum_q\langle \sigma v\rangle_{q\bar{q}}= \langle \sigma v\rangle_{AA}$, and (iii) scenario {\bf s3}:  mainly due to the annihilation into on-shell mediators, assuming that $\langle \sigma v\rangle_{AA}=20\sum_q \langle \sigma v\rangle_{q\bar{q}}$.

\section{Description for experimental constraints}\label{description_expt}

\subsection{Indirect searches}\label{sec:indirect}

\subsubsection{Galactic Center Gamma-Ray Excess}\label{subsec:GCE}

The {\em differential flux of a gamma ray} from a given angular region $\Delta\Omega$, originating from the annihilation of scalar DM particles, is
\begin{eqnarray} 
\label{eq:gammaflux}
\frac{d \Phi_\gamma}{dE} = \frac{1}{\eta}\frac{1}{4\pi m_\phi^2}  \bar{J_\Omega} 
\sum _f\langle \sigma v\rangle_f \frac{dN_\gamma^f}{dE} ,
\end{eqnarray}
where $\eta \equiv$ ``2" is for the self-conjugated DM ({\it e.g.} real scalar DM) and ``4" for non-self-conjugated DM ({\it e.g.} complex scalar DM), $dN_\gamma^f/dE$ is the energy spectrum of DM prompt gamma rays produced  per annihilation into the final state {\em f}, and the scalar DM mass is denoted by $m_\phi$. 
The factor $\bar{J}_\Omega$ is an average of the $J$ factor over the solid angle $\Delta\Omega$, covering the region of interest (ROI), given by
\begin{eqnarray} 
\label{eq:J}
\bar{J_\Omega} =\frac{1}{\Delta\Omega} \underbrace{ \int_{\Delta\Omega}  \int_{\rm l.o.s.} ds \rho^2(r(s,\psi))d\Omega }_J 
\equiv
\mathcal{J}_\gamma \cdot  \left( \frac{\rho_\odot}{0.4~\text{GeV/cm}^3} \right)^2
\cdot
\bar{J}_\Omega^{\mathrm{c}} 
\,,
\end{eqnarray}
where the integral is performed along the line of sight (l.o.s.), $\rho(r)$ is the DM halo profile with $r$ being the distance from the Galactic center, and $\psi$ is the angle observed away from the Galactic center and $d\Omega\equiv \cos{b}\ d\ell\ db$ satisfying $\cos\psi=\cos{b}\cdot \cos{\ell}$ with $\ell$ and $b$ being the longitude and latitude in the Galactic polar coordinate, respectively.
$\bar{J}_\Omega^{\mathrm{c}}$ is the canonical value of $\bar{J}_\Omega$, while $\mathcal{J}_\gamma$ parametrizes the deviation from the canonical profile due to variation of the DM distribution slope $\gamma$. The values of $\bar{J}_\Omega^{\mathrm{c}}$ and $\mathcal{J}$, sensitive to astrophysical uncertainties, depend on the observational ROI in a particular analysis. Following \cite{Calore:2014xka,Calore:2014nla} where the tail of the spectrum has extended to higher energy, we employ the ROI of $|\ell |<20^\circ$ and $2^\circ<|b|<20^\circ$  (i.e., $40^\circ \times 40^\circ$ square centered on the GC with the latitude $|b|>2^\circ$) to study GC gamma-ray emission.

 We adopt the Galactic DM density distribution described by a generalized Navarro-Frenk-White (gNFW) halo profile  \cite{Navarro:1995iw,Navarro:1996gj},
 \begin{equation}
 \label{eq:gNFW}
 \rho(r)=\displaystyle \rho_{\odot} \left(\frac{r}{r_\odot}\right)^{-\gamma} \left(\frac{1+r/r_s}{1+r_\odot /r_s}\right)^{\gamma-3}.
 \end{equation}
As the canonical values we choose the scale radius $r_s=20$~kpc, the slope  $\gamma=1.26$, and the local DM density $\rho_\odot =0.40$~GeV/cm$^3$ at $r=r_\odot=8.5$~kpc, which is the distance of the Solar System from GC. 
The uncertainty of  the profile near the Galactic center remains large.  For instance,  taking the allowed values, $\gamma \in [1.1,1.36]$ and $\rho_\odot \in [0.2, 0.6]$~GeV/cm$^3$, the resulting uncertainties related to the $J$ factor read 
  $\mathcal{J}_\gamma\in [0.66, 1.3]$ and   $\bar{J}_\Omega / \bar{J}_\Omega^{\mathrm{c}}\in [0.17, 3.0]$.

In the numerical analysis of $\phi \phi \to A^* \to q {\bar q}$,  we use the two-body spectra  $dN_\gamma^q/dE$ from the PPPC4DMID results, which, generated using PYTHIA 8.1 \cite{Sjostrand:2007gs}, have included the electroweak corrections  \cite{Cirelli:2010xx,Ciafaloni:2010ti}.  When the DM annihilation into two on-shell mediators occurs, the process has two final states, i.e., four quarks produced, and the photon spectrum $d{\bar N}_\gamma^q/dE$ defined in the DM center of mass frame can be written in terms of the spectrum $(dN_\gamma^q/dE')_A$ described in the rest frame of the mediator by considering a Lorentz boost \cite{Elor:2015tva},
\begin{equation}
\frac{d\bar{N_\gamma^q}}{dE} = \frac{2}{m_\phi} \int^{t_{\rm max}}_{t_{\rm min}} \frac{d x'}{x' \sqrt{1-\epsilon^2}} \Big( \frac{d N_\gamma^q}{dx'} \Big)_A \,,
\end{equation}
where
 \begin{equation}
 t_{\rm max} = {\rm min} \Big[1, \frac{2 x}{\epsilon^2} (1+\sqrt{1-\epsilon^2}) \Big], \quad
  t_{\rm min} = \frac{2 x}{\epsilon^2} (1-\sqrt{1-\epsilon^2}),
   \end{equation}
with $x=E/m_\phi$, $x'=2E'/m_A$, and $\epsilon= m_A/m_\phi$.

We fit the DM mass and its corresponding annihilation cross section $\langle\sigma v\rangle$ to the prompt gamma energy spectrum of the GC excess extracted by Calore, Cholis, and Weniger (CCW) \cite{Calore:2014xka}. CCW studied Fermi-LAT data covering the energy range 300 MeV$-$500 GeV in the inner Galaxy, where the ROI extended to a $40^\circ \times 40^\circ$ square region around the Galactic center with $|b|\leq 2^\circ$ masked out.  
We perform the goodness of fit with a $\chi^2$ test statistic for the total annihilation cross section  $\langle\sigma v\rangle$  and $m_\phi$,

\begin{equation}
\chi^2
=
\left[\frac{d\Phi_\gamma}{dE_i}(m_\phi,\langle\sigma v\rangle) -
\left(\frac{d\Phi_\gamma}{dE_i}\right)_{\text{obs}}\right]
\cdot \Sigma_{ij}^{-1} \cdot
\left[\frac{d\Phi_\gamma}{dE_j}(m_\phi,\langle\sigma v\rangle) -
\left(\frac{d\Phi_\gamma}{dE_j}\right)_{\text{obs}}\right]\,,
\end{equation}
where a total of 24 bins are used in the energy range 300 MeV$-$ 500 GeV,  $ d\Phi_\gamma/dE_i $ and $ (d\Phi_\gamma/dE_i )_\text{obs}$ stand for the model-predicted and observed GCE flux in the $i{\rm th}$ energy bin, respectively. Here the covariance matrix $\Sigma$ contains  statistical error, empirical model systematics and residual systematics, where the latter two are correlated across different energy bins.

\subsubsection{Null measurements of gamma-ray emission from  dwarf spheroidal galaxies}\label{subsec:dSph}

  The dSphs with little dust and gas are DM dominated. Most of them are expected to have no known astrophysical gamma-ray sources.
The Fermi-LAT Collaboration has recently presented a binned Poisson maximum-likelihood analysis on the gamma-ray flux from a large number of Milky Way dSphs based on 6 years of data  \cite{Ackermann:2015zua,Fermi-LAT:2016uux}.  The Fermi-LAT data contain 24 bins, and the bin energy range spans from 500~MeV to 500~GeV, and no significant gamma-ray excess was measured from the dSphs. The observation can offer stringent constraints on the annihilation cross section of DM particles.
We will use combined results of the 15 dSphs, recently reported by Fermi-LAT \cite{Ackermann:2015zua}, in the theoretical analysis.     Using the joint likelihood method, we compute and then add the bin-by-bin delta-log-likelihood  for the 15 dSphs. 
The profile likelihood ratio test statistic (TS), following a $\chi^2$ distribution, is given by
\begin{equation}
\text{TS}=-2 \sum_{k=1}^{N_{\rm dSph}} 
\ln \left[ \frac{ \mathcal{L}_{k} ( \langle \sigma v\rangle_0,  \hat{J}_k ; m_\phi | {\rm data}) }{ \mathcal{L}_{k} ( \overline{\langle \sigma v\rangle}, \bar{J}_k ; m_\phi | {\rm data}) }\right]  \,,
\end{equation}
with the profile likelihood for target (each dSph) $k$ described as
\begin{eqnarray}
 \mathcal{L}_{k} ( \langle \sigma v\rangle,  J_k ; m_\phi | {\rm data}) 
 =
 \left(
 \sum_{i=1}^{N_{\rm bin}}  
 \mathcal{L}_{ki} ( \langle \sigma v\rangle,  J_k ; m_\phi | {\rm data}) 
 \right)
 \cdot  \mathcal{L}_{J_k}  \,,
 \end{eqnarray}
where $N_{\rm dSph}=15$ and $N_{\rm bin}=24$ are the numbers of dSphs and bins, respectively. Here the binned likelihood, $ \mathcal{L}_{ki}$, is a function of the gamma-ray's energy flux within the ``$i$th" bin\footnote{The data is available from the website: http://www-glast.stanford.edu/pub\_data/1048/}, $E_i \Delta\Phi_\gamma$, and
 the $J$ factor likelihood for a target $k$ is modeled by a normal distribution \cite{Lindholm:2015thesis},
\begin{equation}\label{eq:L_jfactor}
\begin{aligned}
  \mathcal{L}_{J_k}  = \frac{1}{\ln(10) J_{{\rm o},k}
    \sqrt{2 \pi} \sigma_k} e^{-\left(\log_{10} J_k- \log_{10} J_{{\rm o},k} \right)^2/(2\sigma_k^2)} \,.
\end{aligned}
\end{equation}
For the $J$ factor of a target $k$, $J_k$ is its expected value, and $ J_{{\rm o},k}$ is the measured nominal value of an error $\sigma_k$.
 For a given $m_\phi$, $\overline{\langle \sigma v\rangle}$ and $\bar{J}_k$ are the respective best-fit values for the DM annihilation cross section and the $J$ factor, corresponding to the minimum value of $ -2\sum_{k=1}^{k= N_{\rm dSph}} \ln\mathcal{L}_{k}$, while
  $\hat{J}_k$ are the conditional maximum likelihood estimators (MLEs) of the nuisance parameters when  $\langle \sigma v\rangle$ is fixed to a given value. $\langle \sigma v\rangle_0$ is the upper limit of the cross section, corresponding to the null measurement, and its 95\% confidence level (C.L.) limit can be obtained by increasing the value of $\langle \sigma v\rangle$ from $\overline{\langle \sigma v\rangle}$ until $\text{TS}=2.71$. 
 
 We will use the results for $ J_{{\rm o},k}$ and $\sigma_k$ given in Ref.~\cite{Ackermann:2015zua}. These values were obtained assuming an NFW halo profile and shown to be insensitive to the models of dark matter density profile if the central value of the slope is less than $1.2$. 
On the other hand, we also consistently use the PPPC4DMID results to generate the relevant two-body and four-body $dN_\gamma^f/dE$ spectra as the studies of the GC gamma-ray excess.

\subsection{Direct detections}\label{sec:direct}

\subsubsection{The effective Lagrangian at the nucleon level}

To obtain the differential DM-nucleus scattering rate at direct detection experiments,  we rewrite the pseudoscalar-quark interacting Lagrangian at the nucleon level  with the replacement,
 \begin{eqnarray}\label{eq:EFFlagrangian}
 i A \sum_q g_q \bar q  \gamma^5 q  \rightarrow  i A \sum_{N=p,n}  c_N \bar N  \gamma^5 N \,,
\end{eqnarray}  
where the pseudoscalar coupling with the nucleon (labeled by p $\equiv$ proton and n $\equiv$ neutron) can be expressed in terms of quark spin contents of the nucleon  \cite{Cheng:2012qr}, $\Delta q^{(N)}$, and is given by
\begin{equation}
c_N=\sum_{q=u,d,s}  \frac{m_N}{m_q}  \left( g_q  - \sum_{q'=u,\dots,t} g_{q'} \frac{\bar m}{m_{q'}} \right) \Delta q^{(N)} \,,
\end{equation}
with $\bar{m} =(1/m_u +1/m_d +1/m_s)^{-1}$. The ratio $c_p/c_n$ is sensitive  not only to $m_u/m_d$, but also to the values of  $\Delta q^{(N)}$'s \cite{Yang:2016wrl}. For illustration, numerically we will adopt the following two sets of  parameters \cite{Cheng:2012qr}:
\begin{eqnarray}\label{eq:set1}
\text{\bf set 1}:\quad 
\Delta u^{(p)} = \Delta d^{(n)} =+0.84 \,, \ \
\Delta d^{(p)} = \Delta u^{(n)} =-0.44 \,, \ \
\Delta s^{(p)} = \Delta s^{(n)} =-0.03 \,,
\label{eq:Deltaq}
\end{eqnarray}
\begin{eqnarray}\label{eq:set2}
\text{\bf set 2}:\quad 
\Delta u^{(p)} = \Delta d^{(n)} =+0.85 \,,\  \
\Delta d^{(p)} = \Delta u^{(n)} =-0.42 \,,\  \
\Delta s^{(p)} = \Delta s^{(n)} =-0.08 \,,
\label{eq:Deltaq-2}
\end{eqnarray}
Meanwhile, we will use $m_u/m_d=0.48$ for set 1 and $m_u/m_d=0.59$ for set 2, where the ratios are consistent with $m_u/m_d=0.48\pm 0.10$ given in PDG \cite{PDG}. In addition to that,  all quark masses are also used from PDG  and consistently rescaled to $\mu=1$~GeV in the direct search studies.

\subsubsection{The nuclear recoil rate and DM velocity distribution function}

In direct detection, the scattering rate of the DM particles off target nuclei is given by
\begin{eqnarray}\label{eq:rate}
\frac{dR_T}{dE_R}=N_T\frac{\rho_\odot}{m_{\phi}}\int_{v_{\text{min}}(E_R)}  v f_{\oplus}(\vec{v},t)\frac{d\sigma_T}{dE_R}   d^3v \,,
\end{eqnarray}
where  $N_T$ is the atomic number of the target, $f_\oplus(\vec{v},t)$ is the DM velocity distribution in the Earth frame,  and $v_{\min}$ is the minimal DM velocity needed for a nucleus to scatter with a recoil energy $E_R$.
Throughout this paper, I will consistently use the  local DM density $\rho_\odot \simeq 0.4$ GeV/cm$^3$.  
Here, $f_\oplus(\vec{v},t)$ can be rewritten in terms of its Galactic frame distribution, $\tilde f(\vec{v})$, 
\begin{eqnarray}
f_\oplus(\vec{v},t)=\tilde f(\vec{v}+\vec{v}_\oplus(t)), 
\end{eqnarray}
where $\vec{v}_\oplus$ is the relative velocity of the Earth in this (Galactic) frame, and its magnitude can be approximated by
\begin{equation}
v_\oplus(t) \simeq \left[v_\odot +u_E \cos\gamma \cos \left( 2\pi \frac{t-152.5\ \text{days}}{365.25\ \text{days}}\right)\right] \, \text{km/s} \,,
\end{equation}
with  $v_\odot\simeq 232~ \text{km/s}$  being due to the motion of the Sun relative to the Galactic frame, $u_E\simeq 30~\text{km/s}$ being  the relative speed between the Earth and Sun,  and $\gamma\simeq 60^\circ$ being the angle  between the Milky Way's disk and Solar System \cite{Belli:2002yt,Schoenrich:2009bx,Freese:2012xd}.

The gNFW halo profile given in Eq.~(\ref{eq:gNFW}) exhibits a double-power law density; the inner log slope of the halo density near the core is $-\gamma$, while the log slope at large radii is $-3$. However, the isotropic Maxwellian velocity distribution, which is usually used in the direct detection analysis, arises from the density slope of $-2$.   To have a velocity distribution function consistent with the gNFW halo profile given in Eq.~(\ref{eq:gNFW}), we adopt the isotropic velocity distribution ansatz, which can reproduce the Eddington formula with double power-law density,  given by \cite{Lisanti:2010qx},
\begin{eqnarray}
\tilde f_{MB}(\vec{v};v_0,v_{\text{esc}})\propto  \left[ \exp\left(\frac{v_{\rm esc}-v^2}{k v^2_0}-1\right) \right]^k \Theta(v_{\text{esc}}-v) ,
\end{eqnarray} 
where $v_0$ is the dispersion, $v_{\rm esc}$ is the escape velocity, and $k\simeq 2$,  the best fit to the gNFW profile model,  is controlled by the outer slope of the halo density. We use $v_0 = 220$ km/s and  $v_{\text{esc}} = 544$ km/s \cite{Smith:2006ym}.

\subsubsection{The differential rate}

At the leading order, the relevant nucleon matrix element is related to the following nonrelativistic operator:
\begin{eqnarray}
\langle \phi(p'), \, N(k')| {\cal O}^{\rm N} |\phi(p),\, N(k) \rangle  &\to& -2 {\cal O} _{10}=  -2 i \vec{q}\cdot\vec{S}_N\,,
\end{eqnarray} 
where the momentum transfer is $\vec{q}=\vec{p}^{\, \prime}-\vec{p}$, $\vec{S}_N$ is the nucleon spin, and $O^{\rm N}$ is given by
\begin{eqnarray}
&&  {\cal O}^{\rm  N} = \left\{
      \begin{array}{ll}
&\:  \phi^* \phi \, \bar{N} i\gamma_5  N, \quad \:\text{ for the complex scalar DM case},\\
 & \frac{1}{2} \phi^2 \, (\bar{N} i\gamma_5  N), \quad \text{for the real scalar DM case}.\\
      \end{array}
    \right.
\end{eqnarray}
The differential rate can be expressed as
\begin{equation}
\label{eq:Rate-1}
\frac{\mathrm{d} R_T}{\mathrm{d} E_R} =
N_T \frac{\rho_\odot}{m_\phi} \frac{1}{32 \pi } \frac{m_T}{ m_N^2}\,  
 \frac{\vec{q}^2  g_{\phi}^2}{(\vec{q}^2 + m_A^2)^2} 
\sum_{N, N' = p, n} c_N  \, c_{N'} \, \mathcal{F}_{\Sigma''}^{(N, N')}(y, T,t) \ ,
\end{equation}
where
\begin{eqnarray}
\mathcal{F}_{\Sigma''}^{(N, N')}(y, T,t) 
&\equiv& 
\int_{v_{\text{min}}(E_R)} \hspace{-.50cm} \mathrm{d} ^3 v \, \frac{1}{v} \, f_\oplus(\vec{v},t) \, F_{\Sigma''}^{(N, N')}(y, T) \ ,   \\
 \sum_{N,N'=p,n} c_N \, c_{N'} \, F_{\Sigma''}^{N,N'} 
 & \equiv & 
  \frac{4 m_N^2}{\vec{q}^2 m_T^2} \frac{1}{2j+1} 
 \sum_{\text{spin}}  \sum_{N,N'=p,n}   c_N c_{N'}  \left| \langle  T' |  O_{10}  N^+ N^- | T  \rangle \right|^2 \,,
 \end{eqnarray}
with $j$ being the nuclear spin, $``T"$ denoting the target nucleus,  and $N^+$ and $N^-$ being nucleon's nonrelativistic fields involving only creation and  annihilation operators, respectively.  The nuclear form factors $F_{\Sigma''}^{N,N'}$ for various nuclei, given in Refs.~\cite{Fitzpatrick:2012ix,Anand:2013yka}, are functions of  $y = (|\vec{q}|b/2)^2$, where  $b\simeq [41.467/(45A^{-1/3}-25A^{-2/3})]^{1/2}$ is the harmonic oscillator parameter with $A$ the mass number.

In $|\vec{q}| \to 0$ limit,  the form factors $F_{\Sigma''}^{N,N'}$  relevant to the longitudinal component of the nucleon spin, with respect to the direction of the momentum transfer, are given in the following approximation:
\begin{eqnarray}
  \sum_{N,N'=p,n} c_N \, c_{N'} \, F_{\Sigma''}^{N,N'}(0)
 & \approx &\frac{4}{3}\frac{J+1}{J}\left( c_p \langle S_p \rangle + c_n \langle S_n \rangle \right)^2\,.
 \label{eq:FF-physics}
\end{eqnarray}
The nuclear shell model calculation showed that the expectation values of the nucleon $\langle S_N \rangle$ and the spin of the initial target nucleus $J$ are mainly due to the unpaired nucleon  \cite{Lewin:1995rx}. 
Considering  the current PICO and LUX experiments that will be analyzed in this paper, we can expect that only nuclides  with ground-state spins  $\geq 1/2$ [$^{19}\text{F}(1/2)$,  $^{127}\text{I} (5/2)$, $^{129}\text{Xe}(1/2)$, and $^{131}\text{Xe}(3/2)$] dominantly contribute to $F_{\Sigma''}^{N,N'}$.

\subsubsection{Null results in direct-detection experiments:  LUX and PICO}

To determine the stringent exclusion bounds on physical parameters due to the null results obtained in direct detection experiments, we use a Poisson likelihood function to model the distribution of the observed events and adopt the likelihood ratio test statistic  \cite{Aprile:2011hx,PDG,Baker:1983tu,Hamann:2007pi},
\begin{equation}\label{eq:TS-dd-def}
\text{TS}=-2\lambda ({\bf n}) = -2 \ln [\mathcal{L}(\mathbf{n};\mathbf{n}^{\rm obs}) /  \mathcal{L}(\hat{\mathbf{n}};\mathbf{n}^{\rm obs})]  \,,
\end{equation}
where the likelihood of data is given by the product of Poisson distributions,
\begin{equation}\label{eq:Ln-dd}
\mathcal{L}(\mathbf{n};\mathbf{n}^{\rm obs}) = \prod_i \mathcal{L}_i ( n_i^{\rm th},  n_i^b ; n_i^{\rm obs}) 
=\prod_i  \frac{n_i^{n_i^{\rm obs}} }{n_i^{\rm obs} ! }e^{- n_i } \,,
\end{equation}
$\mathbf{n}=(n_1, n_2, \dots)$ with $n_i$  being the total expected number of events in the $i\text{th}$ energy bin or detector module, $\mathbf{n^{\rm obs}}=(n_1^{\rm obs}, n_2^{\rm obs}, \dots)$ with $n_i^{\rm obs}$ the observed number of events, and $\hat{\mathbf{n}}=(\hat{n}_1, \hat{n}_2, \dots)$ being the MLE of ${\bf n}$, such that $ 0 \leq \lambda({\bf n}) \leq 1$ for any $n^{\rm th}_i \geq 0$.  Here $n_i= n_i^{\rm th}+ n_i^b$ with              
$n_i^{\rm th}$ and $n_i^b$ being the event numbers for the theoretical prediction and expected background, respectively.  The TS of goodness of fit has an asymptotical $\chi^2$ distribution and can be rewritten as
 \begin{equation}\label{eq:TS-dd}
{\rm TS} (\Lambda; m_\phi) = 2 \sum_i \left[ n_i^{\rm th}(\Lambda;m_\phi)  +n_i^b - \hat{n}_i  - n^{\text{obs}}_i  \ln \left( \frac{n^\text{th}_k(\Lambda;m_\phi) + n_i^{\rm b}}{\hat{n}_i} \right)   \right] \,,
\end{equation}
 where the last term in Eq.~(\ref{eq:TS-dd}) is zero when $n_i ^{\rm obs}= 0$. We take $\Lambda\equiv m_A/(g_\phi g)$ as the relevant parameter. Thus,  adopting $\text{TS} = 2.71$ yields a one-side 95\% C.L. upper limit  for $\Lambda$ with respect to any given  $m_\phi$.  
In general, for each bin (or each module) $i$, the number of events theoretically expected at a direct detection experiment can be expressed by
\begin{equation}\label{eq:n-th}
n^{\text{th}}_i= \mathcal{E}_i \int_0^\infty dE_R \sum_T \epsilon_T^i (E_R) \frac{dR_T}{dE_R} \,,
\end{equation}
where  $\mathcal{E}_i$ is the exposure of the experiment and $\epsilon_T^i(E_R)$ is the efficiency and acceptance that a nucleus $T$ with recoil energy $E_R$ is detected.
Considering the background comes with uncertainty  in the form $n_i^b = \bar{n}_i^b \pm \sigma_i^b$, the  likelihood function is modified as 
\begin{equation}\label{eq:dd-mLn}
 \mathcal{L}_i ( n_i^{\rm th},  n_i^b ; n_i^{\rm obs})  
 \to  
 \int_0^\infty d n_i^b \mathcal{L}_i ( n_i^{\rm th},  n_i^b ; n_i^{\rm obs}) \frac{1}{\sqrt{2\pi}\sigma_i^b} 
 \exp \left[ -\frac{(n_i^b-\bar{n}_b^b)^2}{2 (\sigma_i^b)^{2}} \right] \,,
 \end{equation}
where the probability density function of $n_i^b$ is modeled as a Gaussian distribution.

In the following, we briefly describe the very recent LUX and PICO data that are relevant to the TS calculation.

The very recent LUX data, using a xenon target,  released is based on a complete run of  $3.35\times 10^4$ kg$\cdot$days exposure, called WS2014-16 \cite{Akerib:2016vxi}.   The events passing the cut with distance to the wall larger than 4 cm are selected, but those with 3 cm$<r<4$ cm are neglected \cite{daSilva:2017swg}, where the fiducial boundary is defined as 3 cm inwards from the observed wall in S2 space, and the radius is about 6$-$19~cm corresponding to the drifted electrons' drift time between 40 $\mu s$ and 300 $\mu s$ \cite{daSilva:2017swg,lux:2016talk}.
There are about  85\% of events selected, if the number of events is roughly proportional to the fiducial volume. 
Assuming that the DM events distribute evenly below and above  the red solid curve in Fig.~1 of Ref.~\cite{Akerib:2016vxi}, we restrict ourselves to the signal region only below this curve. We therefore multiply the efficiency by an additional factor $1/2 \times 0.85$.  
On the other hand, we quote the efficiency from Fig.~2 of Ref.~\cite{Akerib:2016vxi} and take 3 phd $\leq S1 \leq $ 50 phd (with phd $\equiv$ photons detected), such that four events were observed compared with 3.3 background events predicted \cite{daSilva:2017swg,lux:2016talk}, where the latter due to leakage from the electron recoil band are assumed to be equally distributed in S1.

The PICO-60 used a $\text{CF}_3\text{I}$ target within a bubble chamber and took the data at a continuum of Seitz thresholds between 7 and 20 keV.
The efficiency curves are translated at Seitz threshold energy of 13.6 KeV to the relevant thresholds. The final exposure with all cuts is 1335 kg$\cdot$days, where the
sensitivity is reduced by a trial factor of 1.8. We use  the solid lines given in Fig.~4 of Ref.~\cite{Amole:2015pla}  as the bubble nucleation efficiencies for  $\text{C}$, $\text{F}$, and $\text{I}$.   After all the cuts, the expectations are  $0.5\pm 0.2$ single-bubble events from a background neutron, consistent with the zero single bubble event remaining.

The PICO Collaboration has also reported the result with a total exposure after cuts of 1167 kg$\cdot$days at a thermodynamic threshold energy of 3.3 keV using the PICO-60 dark matter detector and the bubble chamber filled with $\text{C}_3 \text{F}_8$~\cite{Amole:2017dex}. The PICO-60 $\text{C}_3 \text{F}_8$ improves the constraints on the DM parameters, compared with PICO-2L run 2 experiment \cite{Amole:2016pye}.
 We use the best fit efficiency curves for  $\text{C}$ and $\text{F}$, as given by the solid lines in Fig.~4 of Ref.~\cite{Amole:2015lsj}.     
 The background is predicted to be $0.25 \pm 0.09$ single bubble events from neutrons, $0.026\pm 0.007$ events from electron recoils,  and $0.055\pm 0.007$ events from the coherent scattering of $^8\text{B}$ solar neutrinos.  No single-scattering nuclear recoil candidates are observed.

\subsection{Monojet searches}

The studies for monojet plus missing transverse energy (MET) are one of the important searches for dark matter at the LHC. Here, we employ the very recent CMS 13 TeV results, corresponding to an integrated luminosity of 12.9 fb$^{-1}$ \cite{Sirunyan:2017hci}.  Using a profile likelihood ratio, we employed the $\text{CL}_s$ method \cite{Junk:1999kv} to calculate the 95\% confidence level (C.L.) upper limit on the $\not\!\!{E}_{\rm T}$  signal events for $ p p \to \phi \phi j$ at the reconstructed level, where the Standard Model background within a bin is modeled as a Gaussian distribution, but the correlations between uncertainties of the background yields across different $\not\!\!{E}_{\rm T}$ bins are neglected.

  To obtain the constraints of parameters in the simplified model from the upper limit  of monojet signals involving the DM missing transverse energy at the trigger (reconstructed) level,
we implement the present model into FeynRules \cite{Alloul:2013bka,Backovic:2015soa} to get a UFO output \cite{Degrande:2011ua}, which is then used in ${\rm MadGraph5\_aMC}@{\rm NLO}$ \cite{Alwall:2014hca} for the simulation of the relevant monojet events. We set the renormalization scale ($\mu_R$) and factorization scale ($\mu_F$) to be $\xi H_T/2$, where $H_T$ is the sum of the missing transverse energy and the transverse momentum of the jet ($j$), and $\xi \in [1/2, 2]$ denotes the scale uncertainties.

We use MadAnalysis 5 to analyze the events of simulations \cite{Dumont:2014tja}.  The next-to-leading order (NLO) NNPDF3.0 \cite{Ball:2014uwa} parton distribution functions (PDFs) with the corresponding $\alpha_s(M_Z)=0.118$ are used to generate the signal events, which are further hadronized by using PYTHIA8 \cite{Sjostrand:2007gs}.  For jet clustering, consistent with the CMS study, we construct the (AK4) jets by employing the anti-$k_T$ algorithm \cite{Cacciari:2008gp} with the distance parameter $R=0.4$, as implemented in FastJet \cite{Cacciari:2011ma}. The selection cuts for  jets at the reconstructed level are $p_T >20$~GeV and $|\eta| <5.0$, while  the leading one in the event is required to have $p_{T, j_1} >100$~GeV and $|\eta_{j_1}| <2.5$.  
We also impose the jet veto that rejects events if the azimuthal separation between $p_T^{\rm miss}$ and the directions of each of the four highest $p_T$ jets with $p_T>30$ GeV is smaller than $0.5$ radians. This criterion has been used by the CMS to suppress the QCD multijet background.
  
 Because the triggers for events with ${\not\!\!{E}}_{\rm T}>300$~GeV become full efficient at the CMS,  we will take into account the missing transverse energy within the range ${\not\!\!{E}}_{\rm T}=$ 350$-$590~GeV in the numerical analysis. Within these bin widths, the 95\% C.L. upper limit for the total number of signal events due to the generation of the DM particles will be used in the analysis.

\section{Constraints on dark matter parameters}\label{sec:constraints}

\subsection{Indirect detections}\label{subsec:indirect}

\begin{figure}[t!]
  \begin{center}
\hskip0.5cm\includegraphics[width=0.345\textwidth]{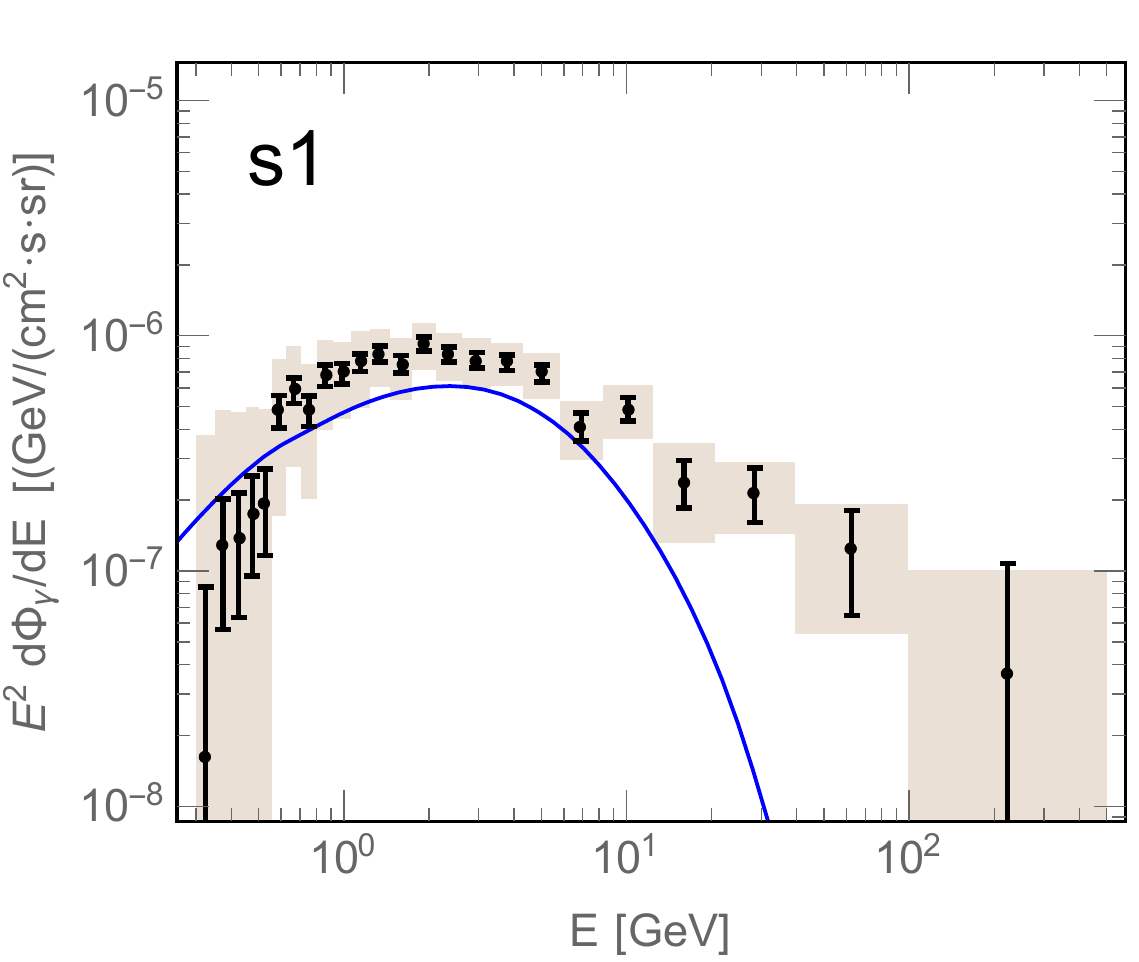}\hskip0.25cm
\includegraphics[width=0.365\textwidth]{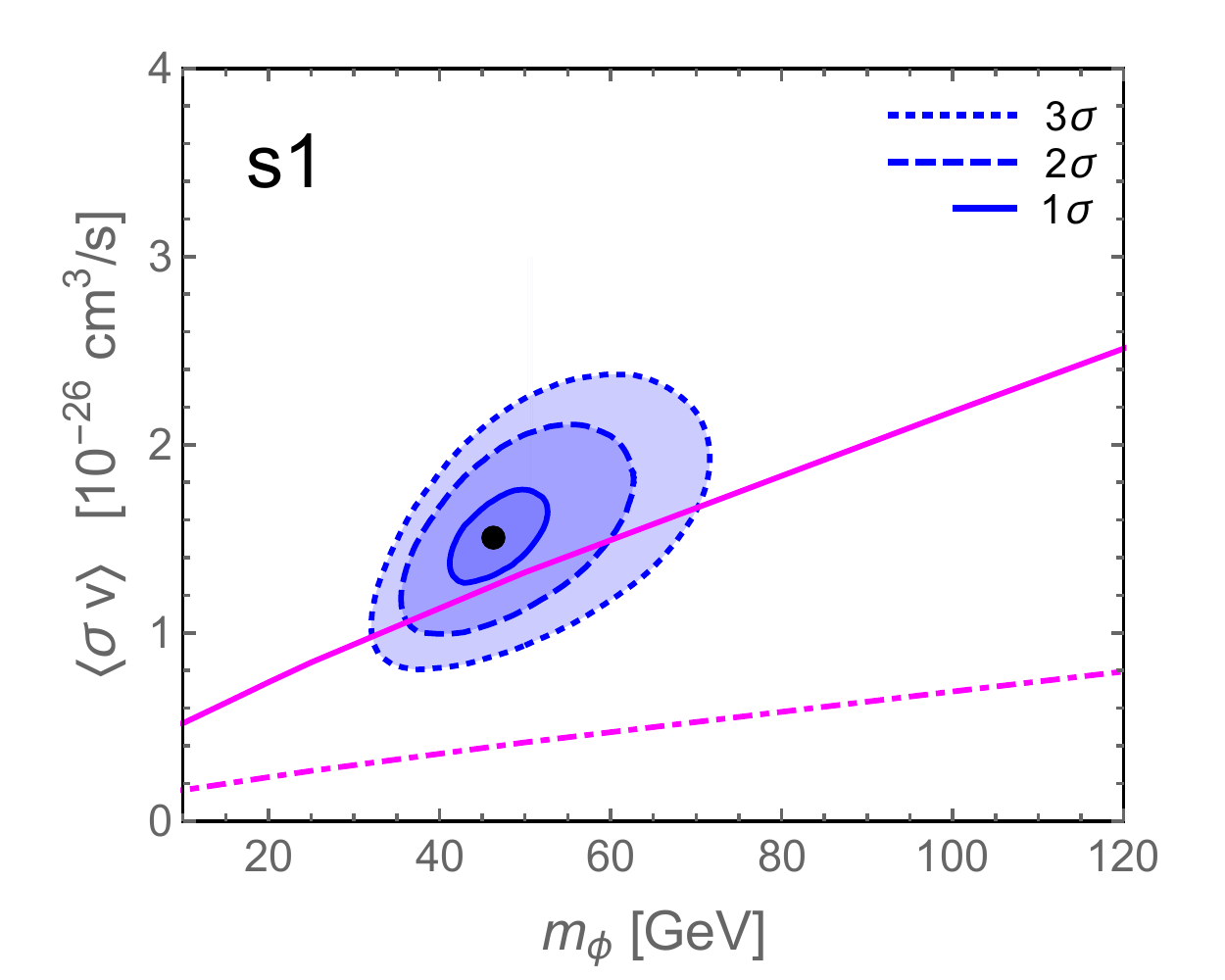} 
\\
\includegraphics[width=0.345\textwidth]{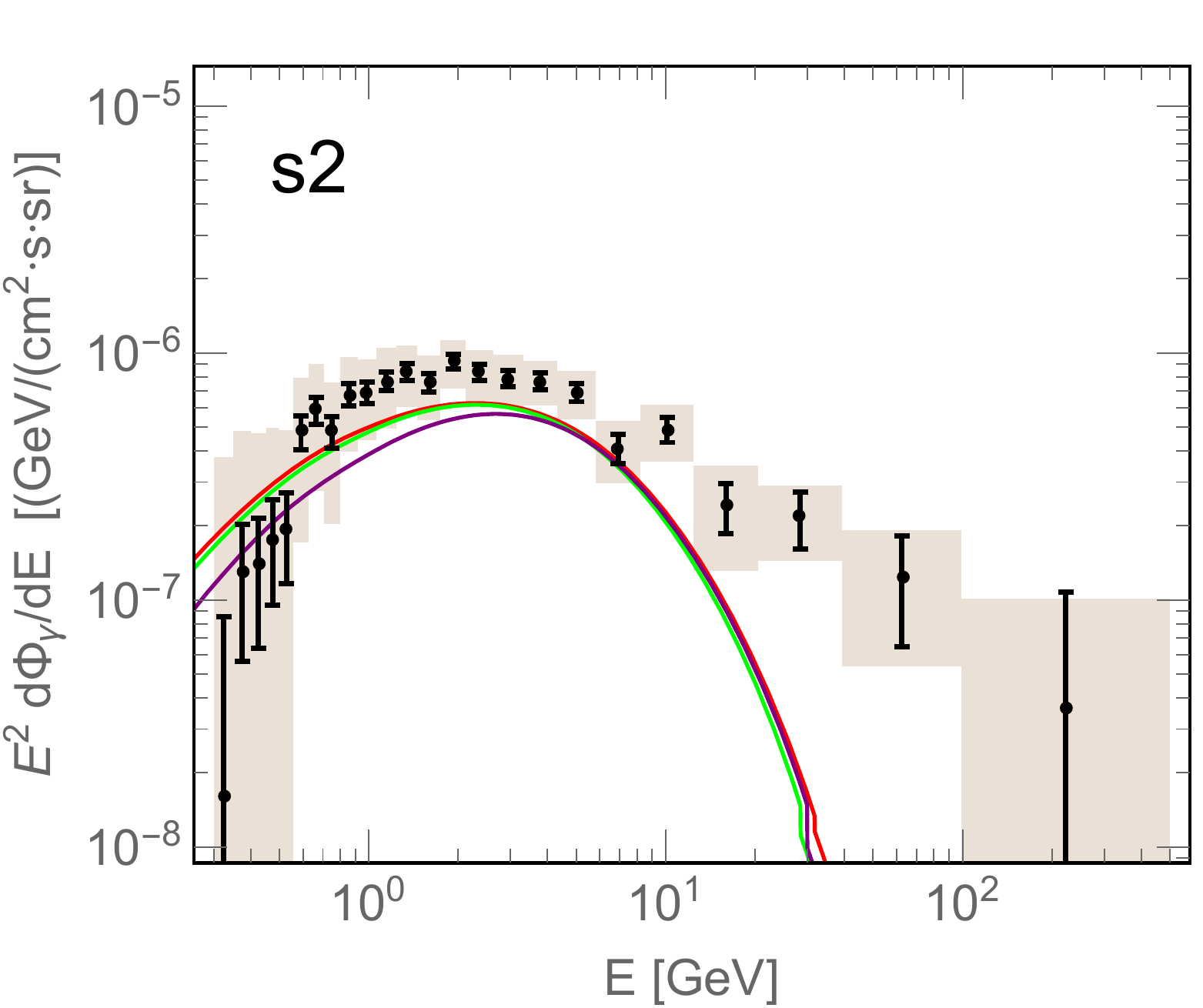}\hskip0.6cm
\includegraphics[width=0.32\textwidth]{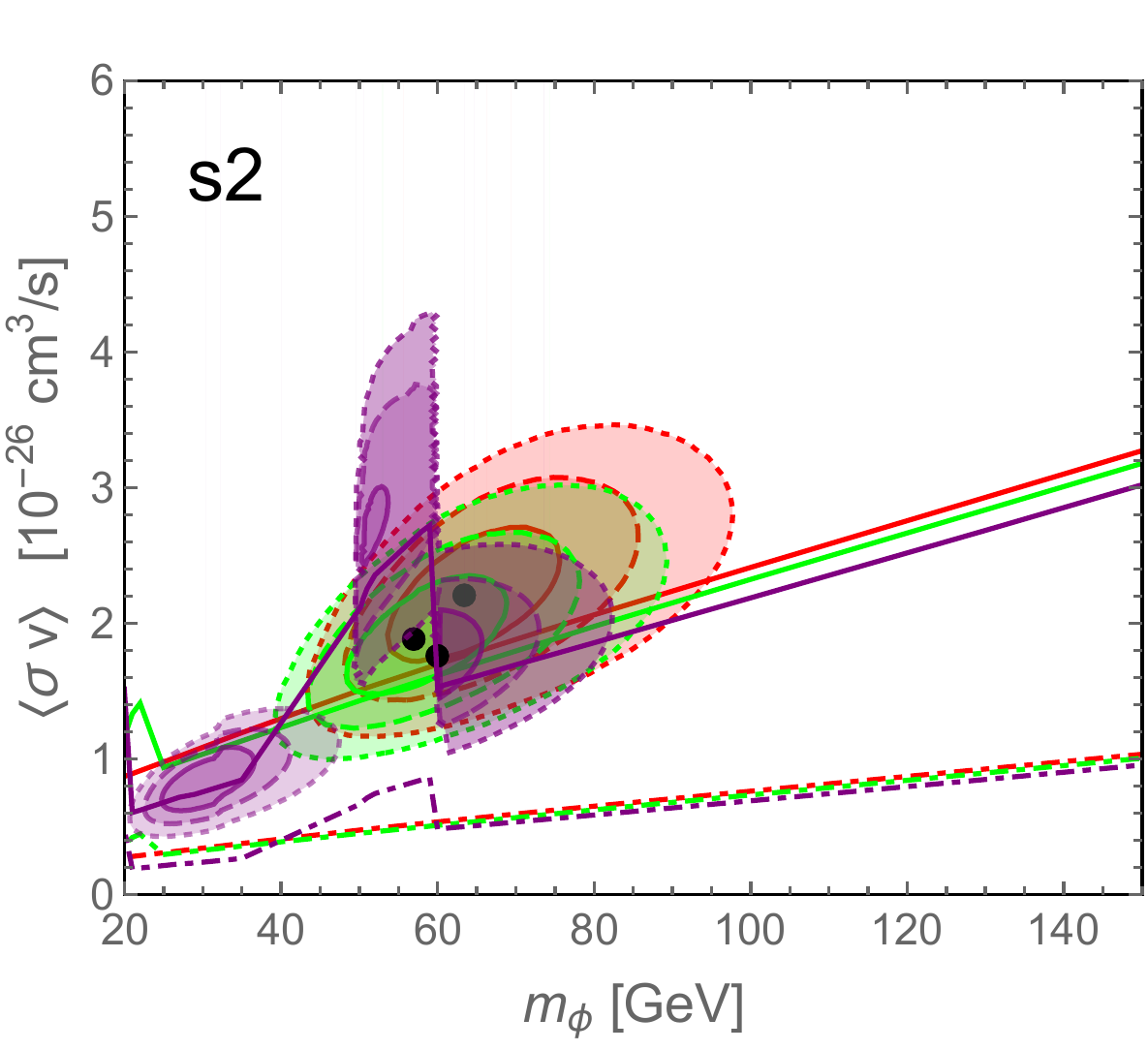} 
\\
\includegraphics[width=0.345\textwidth]{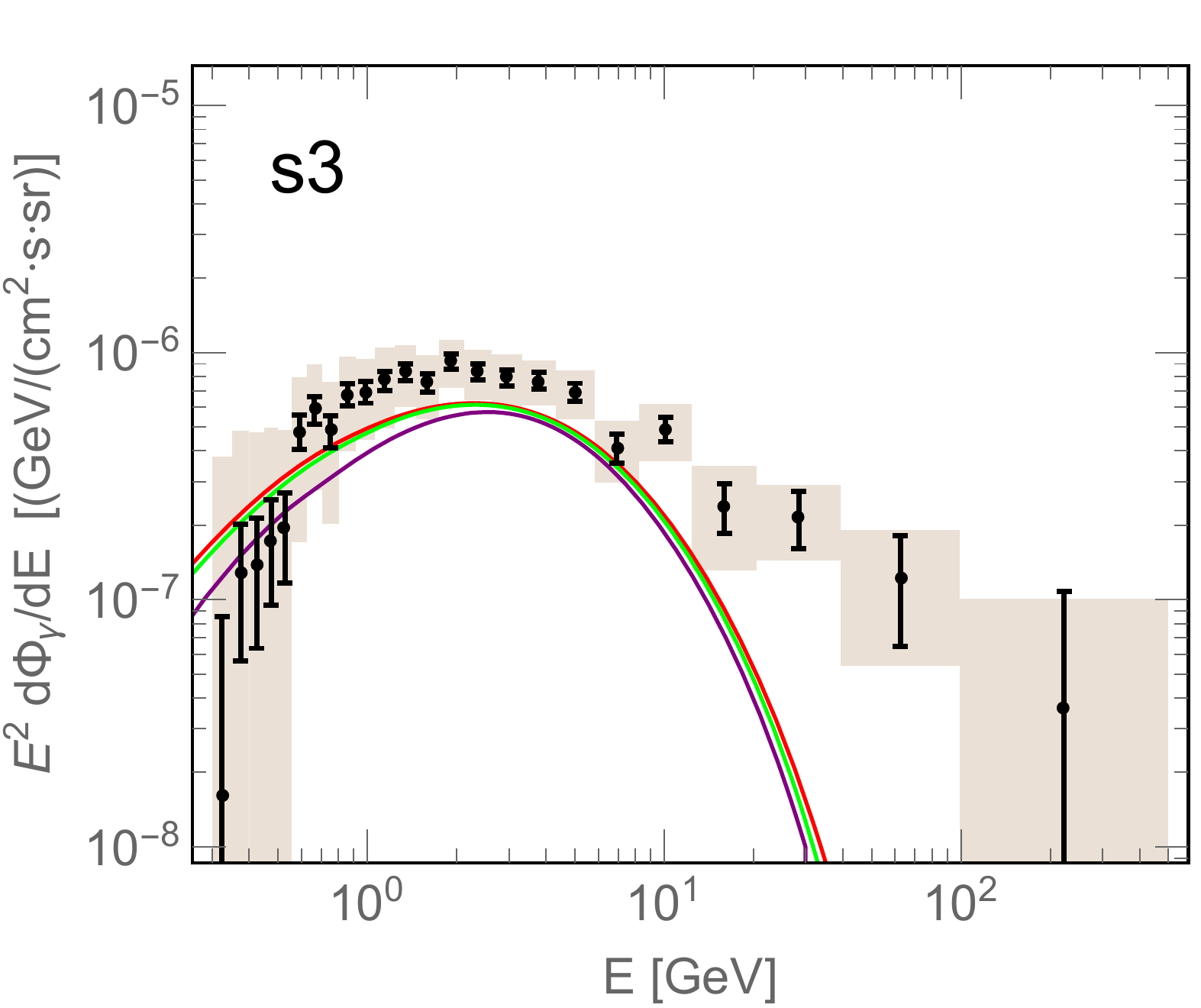}\hskip0.6cm
\includegraphics[width=0.32\textwidth]{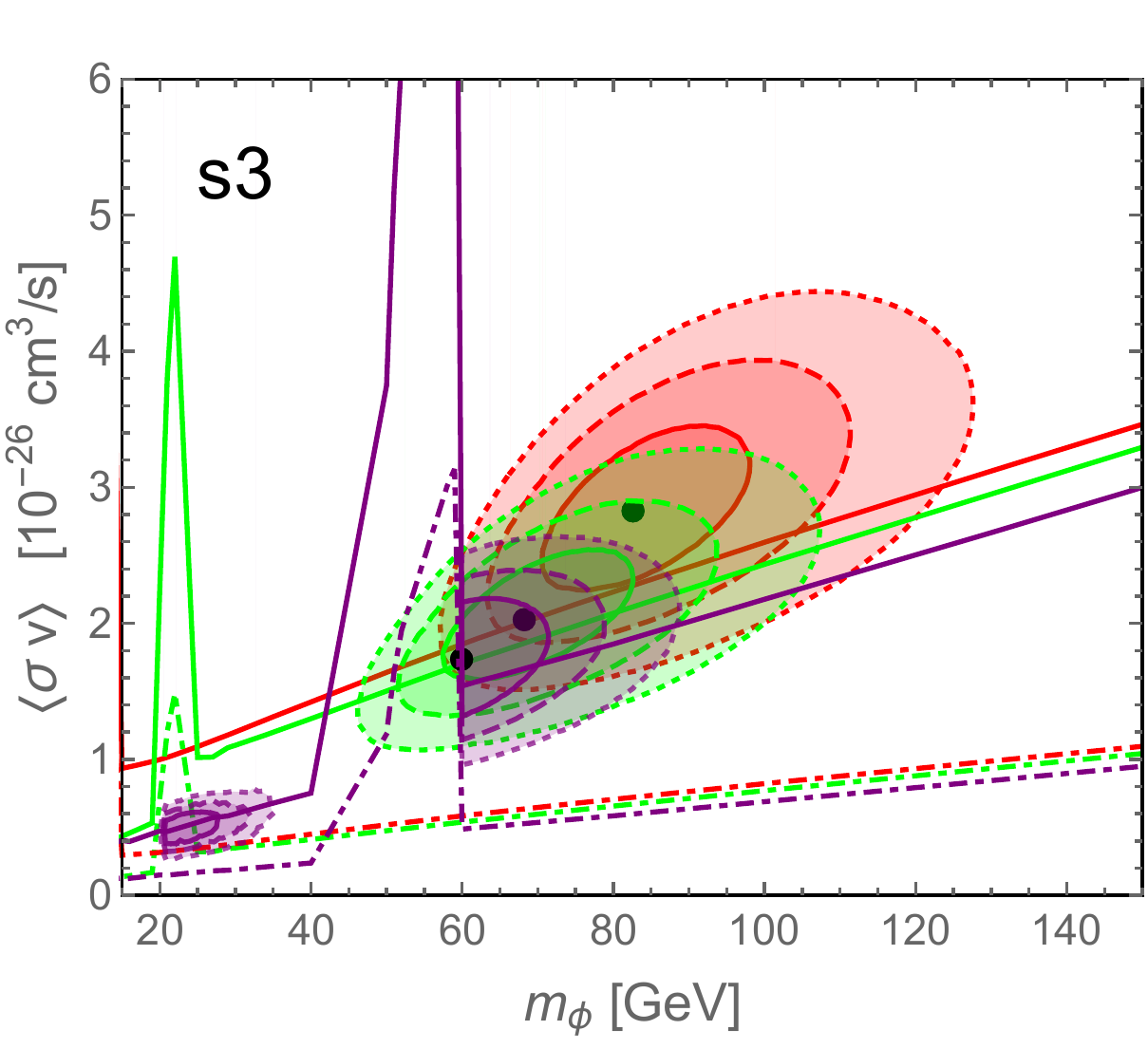} 
\caption{ Fits to prompt GC emission spectra and upper bounds from Fermi-LAT observations of dSphs, where panels from up to down are, respectively, for scenario {\bf s1}, {\bf s2}, and {\bf s3}. For  {\bf s2} and {\bf s3}, the results with $m_A/m_\phi =0.8, 0.5, 0.2$ are denoted as red, green, and purple colors, respectively, and, for {\bf s1}, the GC fitted results are shown in terms of blue color. Left panel: Best fits vs prompt GC emission spectra. The statistical errors are shown by error bars, and systematical errors, including empirical model systematics and residual systematics, are denoted as brown rectangles. Right panel: GC excess regions vs 95\% C.L. upper limits and projected limits from the observations of dSphs, respectively, denoted as the solid and dot-dashed lines (with the same colors in {\bf s1} and {\bf s2} as the corresponding $m_A/m_\phi$ cases).  The GC best-fit value is denoted as the (black) dot. All results refer to $\rho_\odot = \text{0.4 GeV/cm}^3$ and  $\mathcal{J}_\gamma=1$ corresponding to $\gamma=1.26$.}
\label{fig:GC}
\end{center}
\end{figure}

We first fit the DM mass and its corresponding annihilation cross section $\langle\sigma v\rangle$ at an average velocity of $v\sim 10^{-3}c$  to the GC prompt gamma energy spectrum \cite{Calore:2014xka}. The results of fits are plotted in Fig.~\ref{fig:GC}, and summarized in Table \ref{tab:GC}, together with $\pm1\sigma$ errors,  $\chi_\text{min}^2/dof$, and $p$ value.  The results that we show correspond to $\rho_\odot=0.4~\text{GeV/cm}^3$ and $ \mathcal{J}_\gamma=1$ which refers to the inner log slope to be $-1.26 (=-\gamma)$. 
It seems that the model results shown in the left panels of Fig.~\ref{fig:GC} do not appear to be a good fit to the prompt gamma spectrum. However, due to the strong correlations of the systematical errors among different energy bins, the best-fit values give good fits with $p$ values $>0.35$ \footnote{A model with the DM annihilating into $\bar{b} b$ mixed with $\tau^+ \tau^-$ could even have a better fit to the GC gamma-ray excess \cite{Calore:2014nla,Hektor:2015zba}. }.

The DM $s$-channel annihilation is dominated by the ${\bar b} b $ final state for $m_\phi >m_b$, because the mediator interacts with quarks via the Yukawa-like coupling. If the DM annihilation into  the on-shell mediator pair is kinematically allowed, the fitted regions depend on the mass of the mediator, as shown in Fig.~\ref{fig:GC}, where, for illustration, we have shown results for three values of $m_A/m_\phi= 0.2, 0.5$, and $0.8$ in scenarios {\bf s2} and {\bf s3}. 

If the low velocity DM annihilation cross section is dominated by   $\langle \sigma v\rangle_{AA}$ (scenario {\bf s3}), and the produced $A$ particles are nonrelativistic, then  two final states, i.e., four $b$ quarks, are generated in the decays of two $A$ particles, such that the best-fit GC excess results for the DM mass and annihilation cross section are therefore larger by a factor of $\sim 2$ compared to that in the pure $s$-channel annihilation case (scenario {\bf s1}). The best-fit values of $m_\phi$ and the cross section will further decrease either for a smaller $m_A$  due to the fact that the PPPC4DMID gamma-ray spectrum needs to be boosted from the $A$ particle rest frame to the dark matter rest frame in the fit or for a larger contribution arising from the DM $s$-channel annihilation into ${\bar b} b$ (scenario {\bf s2}).

For the scenario {\bf s2}, the parameter region with $m_\phi \sim$ 50$-6$0~GeV and $m_A\sim$~10$-$12~GeV yields a good fit,  for which
the gamma-ray spectrum produced  from the on-shell $A$ decaying into the final state $b {\bar b}$ can be negligible due to the smallness of the phase space, and therefore the result is dominated by the DM $s$-channel annihilation, via an off-shell $A$, into $b {\bar b}$.
On the other hand, for scenarios {\bf s2} and {\bf s3},  the GC gamma-ray excess data can also be fitted by the parameter region  $m_\phi \sim 30$~GeV, 
where in addition to the contribution arising from the DM annihilation into two on-shell mediators, which  dominantly  decay into the ${\bar c} c$ pairs in the final state, {\bf s2} receives a sizable contribution from the DM $s$-channel annihilation into ${\bar b} b$.

The results show that some GC gamma-ray excess allowed regions are excluded by the observations of dSphs. 
It should be noted that the $J$ values of dSphs are obtained subject to the assumptions that the dSphs are spherically symmetric and have negligible binary motions \cite{Lindholm:2015thesis}.
Note also that the uncertainty due to the DM profile of  the Galactic center is not included in the fit of the GC gamma-ray excess; the allowed region of the annihilation cross section shown in Fig.~\ref{fig:GC} could thus be revised by a factor 
$(\bar{J}_\Omega / \bar{J}_\Omega^{\mathrm{c}})^{-1} \in [0.33, 5.88]$. 
One the other hand, the upper limit set by dSphs data is approximately proportional  to the square root of the data size and the square root of the number of observed dSphs \cite{Anderson:2015rox}. Assuming that the Fermi-LAT Collaboration can successfully collect 15-yr gamma-ray emission data  about 60 dSphs \cite{Charles:2016pgz}, the $\langle\sigma v\rangle$ limits will be further improved by a factor of $\sim 3.16$ in the future, as shown by the dot-dashed curves in Fig.~\ref{fig:GC}, and this model is very likely to be testable.

\begin{table}[t]
\setlength{\tabcolsep}{5pt}
\renewcommand{\arraystretch}{1.3}
\center
\begin{tabular}{cccccc} 
\hline \hline 
$m_A/m_\phi$ & $\langle\sigma v\rangle$   & $m_\phi$ &  $\chi^2_\text{min}/dof$  & $p$-\text{value} 
\\[-8pt]
& $[10^{-26}$ cm$^{3}$ s$^{-1}]$   &  [GeV]  &   & \\
\hline
\multicolumn{5}{c}{$\text{\bf s1}: \langle\sigma v\rangle = \sum_q  \langle\sigma v\rangle_{\phi\phi \to \bar{q} q}$}\\
 & $1.51\pm 0.25$ & $46.4_{-5.2}^{+6.3}$ & $1.09$  & $0.35$
\\
\hline
\multicolumn{5}{c}{{$ \text{\bf s2}:  \langle\sigma v\rangle = 2 \sum_q  \langle\sigma v\rangle_{\phi\phi \to \bar{q} q}= 2 \langle\sigma v\rangle_{\phi\phi \to AA}$}}\\
0.8 & $2.20^{+0.51}_{-0.48}$ & $63.5_{-9.9}^{+12.1}$ & $1.06$  & $0.38$
\\
 0.5 & $1.88^{+0.47}_{-0.40}$ & $56.8_{-8.4}^{+12.0}$ & $1.05$  & $0.39$
\\
0.2 & $1.77^{+0.33}_{-0.32}$ & $60.0_{-0.0}^{+5.7}$ & $1.07$  & $0.37$
\\
0.2  & (2.44, 2.96) &  (50.5, 53.5) &
 \\
 0.2  & (0.62, 1.09) &  (24.7, 36.6) &
 \\
\hline
\multicolumn{5}{c}{{$\text{\bf s3}:  \langle\sigma v\rangle = 21 \sum_q \langle\sigma v\rangle_{\phi\phi \to \bar{q} q}= (21/20) \langle\sigma v\rangle_{\phi\phi \to AA}$}}\\
0.8 & $2.82^{+0.63}_{-0.58}$ & $82.8_{-12.2}^{+15.3}$ & $1.07$  & $0.37$
\\
0.5 & $2.23^{+0.90}_{-0.20}$ & $68.2_{-8.0}^{+10.6}$ & $1.07$  & $0.38$
\\
0.2 & $1.74^{+0.44}_{-0.42}$ & $60.0_{-0.1}^{+11.4}$ & $1.13$  & $0.31$
\\
0.2  & (0.38, 0.62) &  (20.6, 27.7) &
 \\
\hline \hline
\end{tabular}
\caption{  \small  Values of spectral fits to the GC gamma-ray emission together with $\pm1\sigma$ errors for three scenarios.  The corresponding $p$ value of $\chi^2_{\rm{min}}$ is given.  $ \langle\sigma v\rangle =\sum_q \langle\sigma v\rangle_{\phi\phi \to \bar{q} q}+\langle\sigma v\rangle_{\phi\phi \to AA}$ is the total cross section, using $\rho_\odot = \text{0.4 GeV/cm}^3$ and  $\mathcal{J}_\gamma=1$ (i.e.,  $\gamma=1.26$).  Results for three values of $m_A/m_\phi= 0.2, 0.5$, and $0.8$ in scenarios {\bf s2} and {\bf s3} are shown, where, for $m_A/m_\phi=0.2$ and $m_\phi\lesssim 60$~GeV, the ranges of $1\sigma$ errors are given in the parentheses.}
\label{tab:GC}
\end{table}

\subsection{Direct detections} 

\begin{figure}[t!]
\begin{center}
\includegraphics[width=0.39\textwidth]{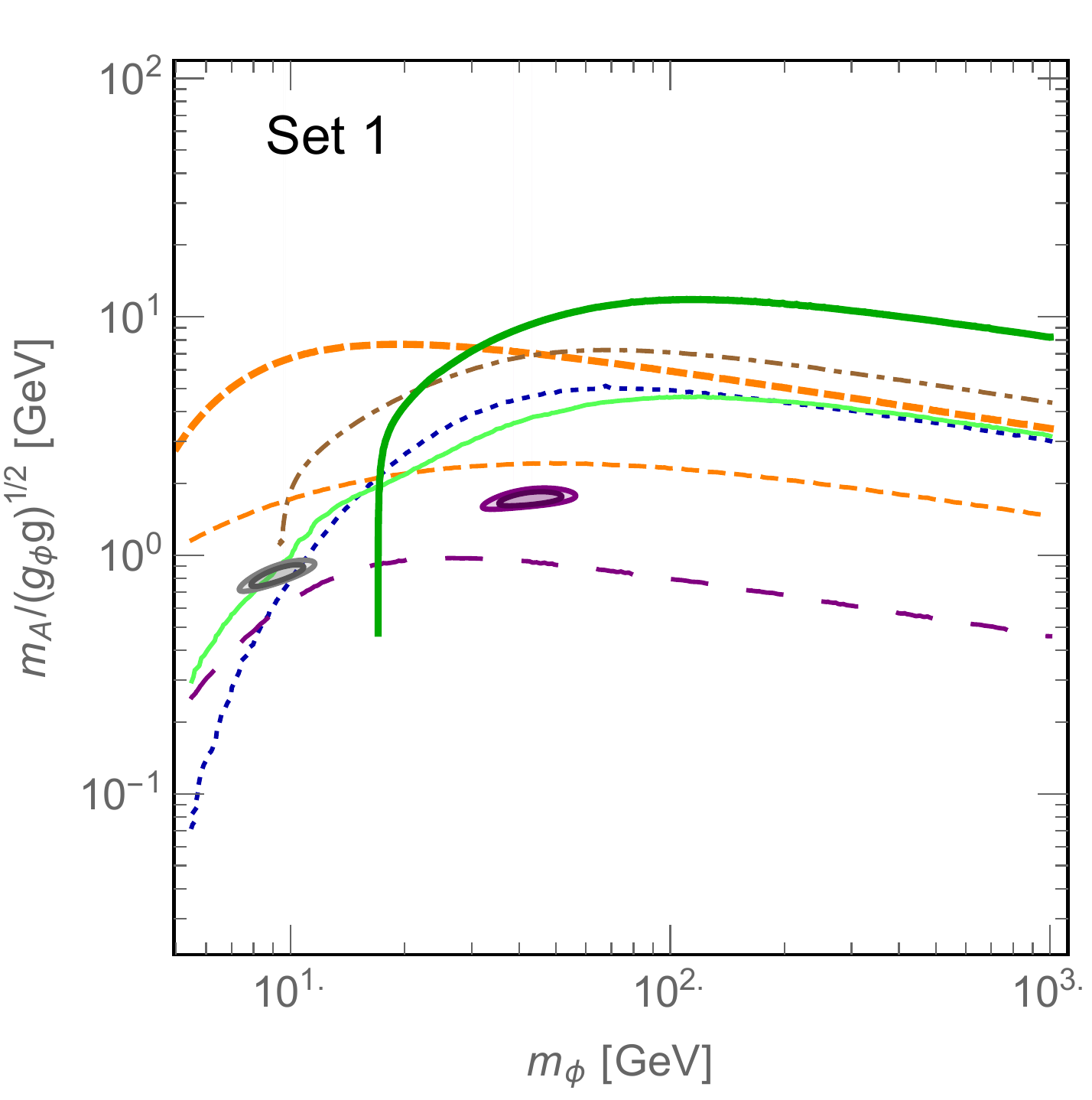}
\hskip0.5cm
\includegraphics[width=0.4\textwidth]{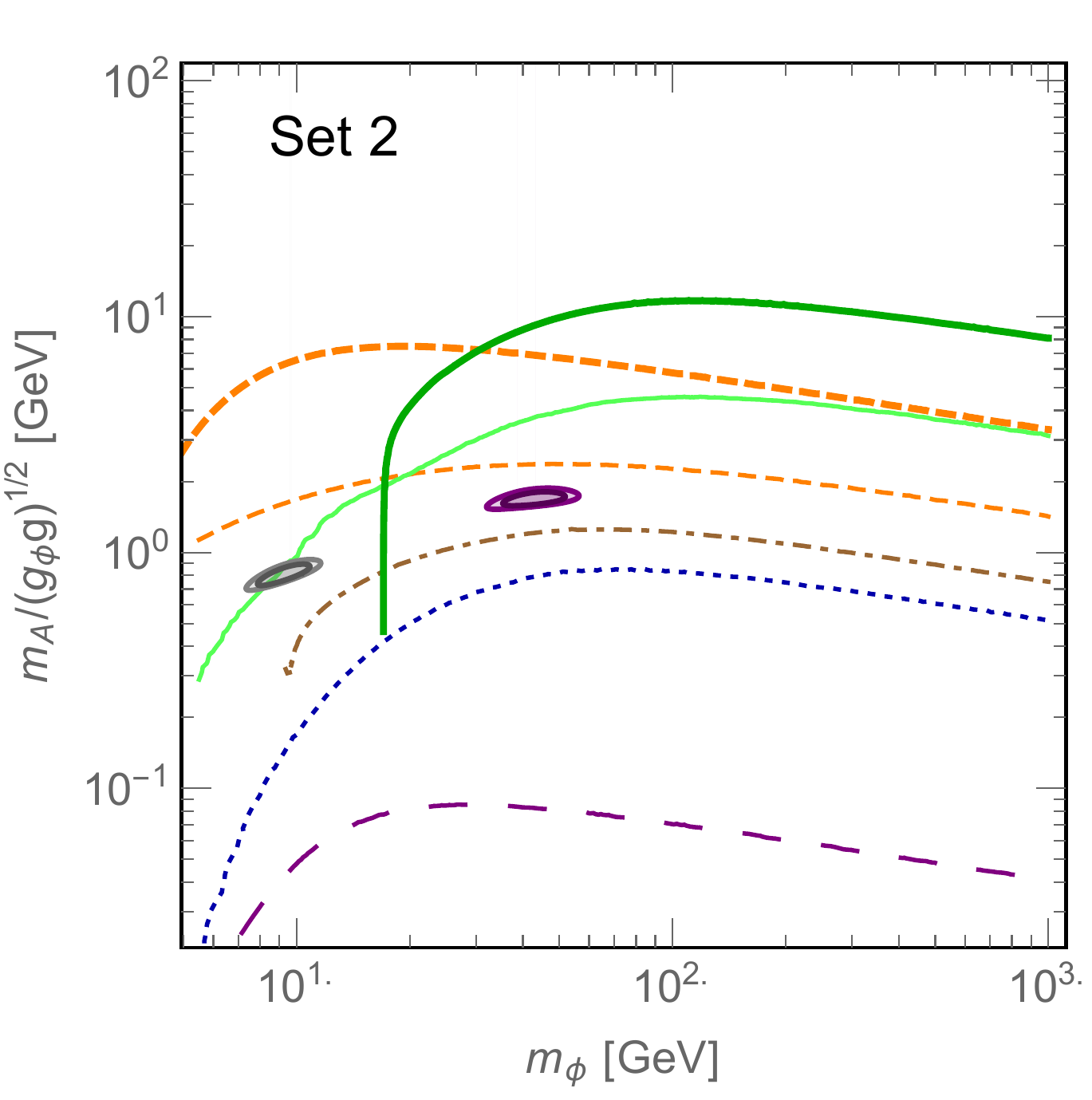}
\caption{95\% C.L. lower limits from PICO-60 ${\rm C}_3 {\rm F}_8$ (thick dashed orange), PICO-60 ${\rm C} {\rm F}_3 {\rm I}$ (thick solid green), and LUX WS2014-16  (dot-dashed brown), together with the DAMA  2$\sigma$ (inner shaded region) and 3$\sigma$ (outer shaded region) allowed regions. For comparison, the exclusion results  from earlier measurements PICASSO (thin dashed orange),  COUPP (thin solid green), XENON100 (dotted blue), SuperCDMS (long-dashed purple) are also shown.  Set 1 parameters for $\Delta q^{(N)}$ and  $m_u/m_d= 0.48$ is used in the left panel, while set 2 and $m_u/m_d= 0.59$ in the right panel.}
\label{fig:DirectDetect}
\end{center}
\end{figure}

 In Fig.~\ref{fig:DirectDetect}, we show  the exclusion bounds from  the null measurements by LUX WS2014-16 and PICO-60 in the $[m_\phi, m_A/(g_{\phi} g)^{1/2}]$ plane. Here, the contact interaction, which is a good approximation for $m_A \gtrsim 300$~MeV, is taken, i.e., $m_A^2\gg \vec{q}^2$.  For comparison, we plot the DAMA modulation \cite{Bernabei:2010mq} $2\sigma$ and $3\sigma$ allowed regions, and exclusion results extracted from earlier measurements by PICASSO \cite{PICASSO}, COUPP \cite{COUPP}, XENON100 \cite{XENON100}, and SuperCDMS \cite{SuperCDMS}, where the method for treating these null data can be referred to Ref.~\cite{Yang:2016wrl}. The approach of the DAMA modulation analysis is similar to that given in Ref.~\cite{Yang:2016wrl}.
For the DAMA signals, two regions at $m_\phi\sim$ 10 GeV  and at $m_\phi\sim$ 40 GeV can be interpreted if the DM particle scatters on the sodium for the former region and iodine for the latter.  However, the PICO measurements seem to strongly disfavor the parameter space fitted from the DAMA modulation data.
  
Our results are summarized as follows. 
(i) The PICO-60 results, mainly due to the unpaired protons in the target nuclei, are insensitive to the choice of the parameter set. 
(ii) If the spin of detector material is mostly due to the unpaired neutron, as LUX (and XENON100) employs Xe (and SuperCDMS uses Ge), the  resulting exclusion limit can be highly suppressed using parameters of set 2. 
(iii) The DAMA results are incompatible with the exclusion bound set by the PICO-60 measurements.  

Although the PandaX-II \cite{Cui:2017nnn} and XENON1T \cite{Aprile:2017iyp}, using also the xenon target,  have recently obtained  
a slightly stronger bound on the (spin-independent) cross section by a factor of 2.5 compared with LUX WS2014-16, the PICO-60 measurements still give the most stringent exclusion bound, which is insensitive to the choice of the parameter set, among the current direct detection experiments\footnote{The spin-independent direct detection cross section can be induced at the one loop. The result, which seems to be reachable in the next generation, was first estimated in Ref.~\cite{Ipek:2014gua} by considering a UV complete fermionic DM model. Because, in the simplified model, the one-loop (box diagram) result is not gauge invariant, we thus do not consider it here. Very recently, Arcadi {\it et al.} \cite{Arcadi:2017wqi} have shown the fermionic DM results that for $m_A\sim 100$~GeV, the sensitivity of the direct detection needs to go beyond the neutrino floor in the simplified model, while it could reach the DARWIN \cite{Aalbers:2016jon} projected sensitivity in a gauge-invariant model.}.  In the following, we will therefore use the PICO-60 results on the model analysis.

\subsection{Monojet  and scenario {\bf s1} }\label{subsec:monojet-result}

The monojet+MET search can provide a relatively stronger constraint on parameters when the mediator is produced on shell at the colliders.  For this case, the monojet cross section can be approximated as $\sigma(pp \to j \phi \phi) \sim \sigma(pp\to j +A) \times {\rm BR}(A\to \phi\phi)$. Therefore, for $m_A>2 m_\phi$,  the monojet search may constrain the GC excess region where only the DM $s$-channel annihilation into the SM quark pair is relevant; in other words, the monojet constraint on the GC excess region can be categorized to the scenario {\bf s1}.  
For the parameter range with $m_A< 2m_t$, the total width of the mediator ($\Gamma_A$) obtained using Eqs.~(\ref{eq:partial-width-1}), (\ref{eq:partial-width-2}), and (\ref{eq:partial-width-3}),  is always small; for instance we have $\Gamma_A/ m_A <0.02$ for $g_\phi, g <5$, consistent with the narrow width approximation.

This monojet constraint depends on the value of $g$. For illustration,  take $m_A> 2 m_\phi$ with $m_A=100~{\rm GeV}$ and $m_\phi=46.4$~GeV as an example. For $g \lesssim 1.3$, the present monojet result cannot provide a sufficient constraint on $g_\phi$ in the range of  $g_\phi \in [0,4\pi] $, because ${\rm BR}(A \to \phi \phi)$ is already larger than $\sim 90\%$ for $g_\phi =1$. On the other hand, for a large $g$ limit, the monojet cross  is proportional to $\sigma(pp \to j \phi \phi) \sim \sigma(pp\to j +A) \times {\rm BR}(A\to \phi\phi) \propto g^2 \times g_\phi^2 /g^2 \propto g_\phi^2$, independent of $g$, such that, for $g \gtrsim 3$, we get that $g_\phi \gtrsim 0.36$ is excluded at 95\% C.L. by the CMS  monojet search.  Numerically,  we find that $g=2$, which will be used in the analysis, can provide a stronger limit on the value of $g g_\phi$.

\begin{figure}[t!]
\begin{center}
\includegraphics[width=0.39\textwidth]{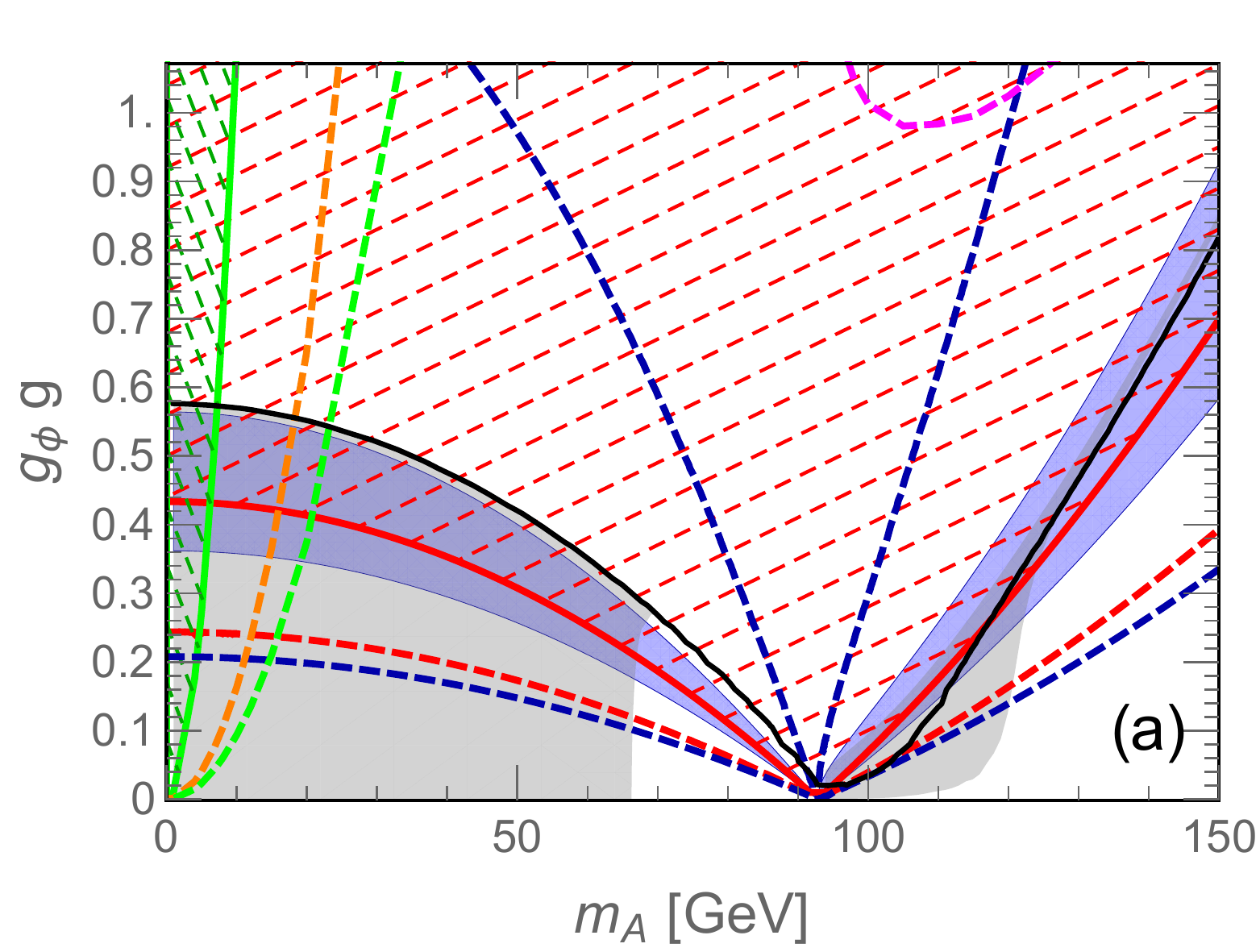}\\
\includegraphics[width=0.39\textwidth]{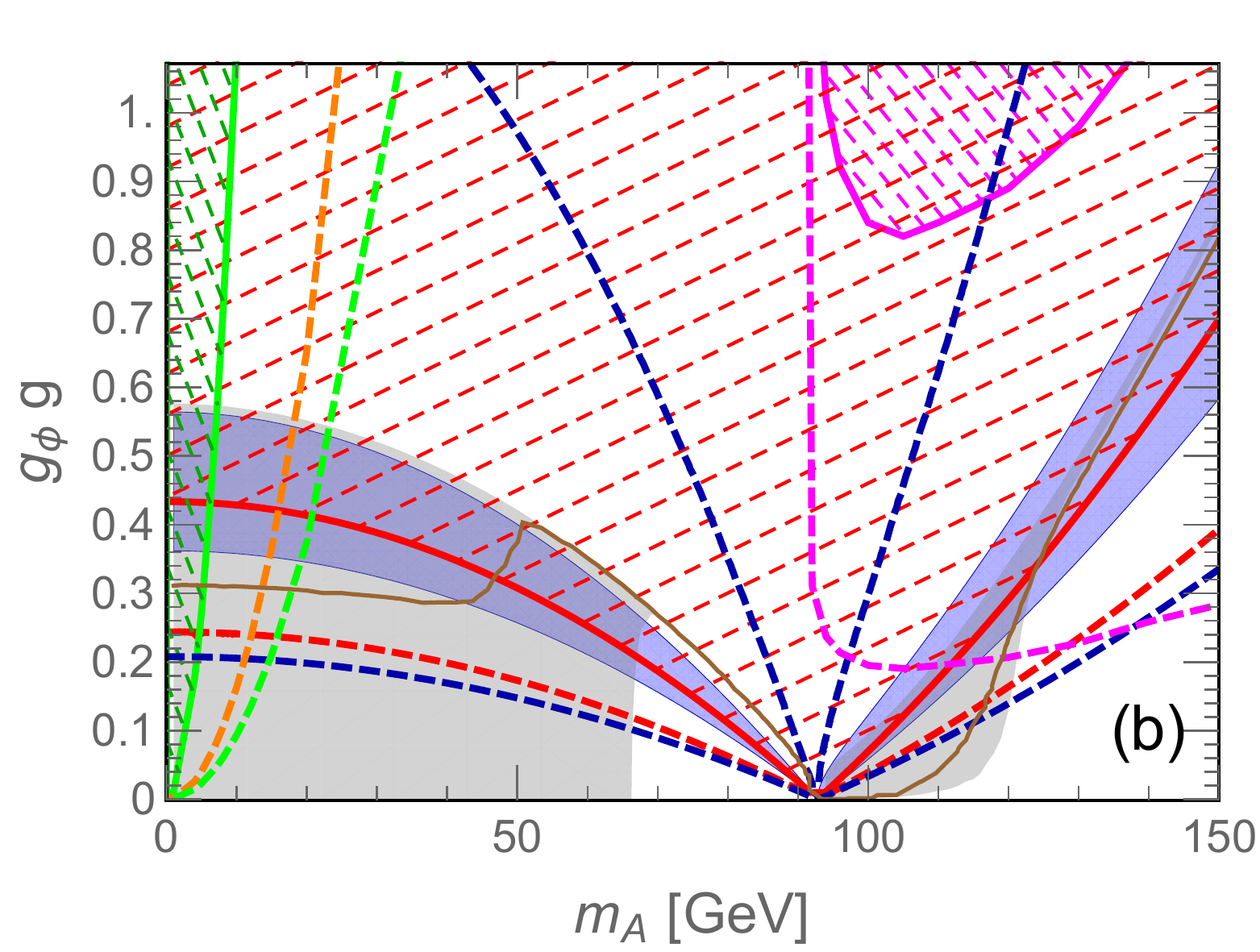} \hskip0.8cm
\includegraphics[width=0.39\textwidth]{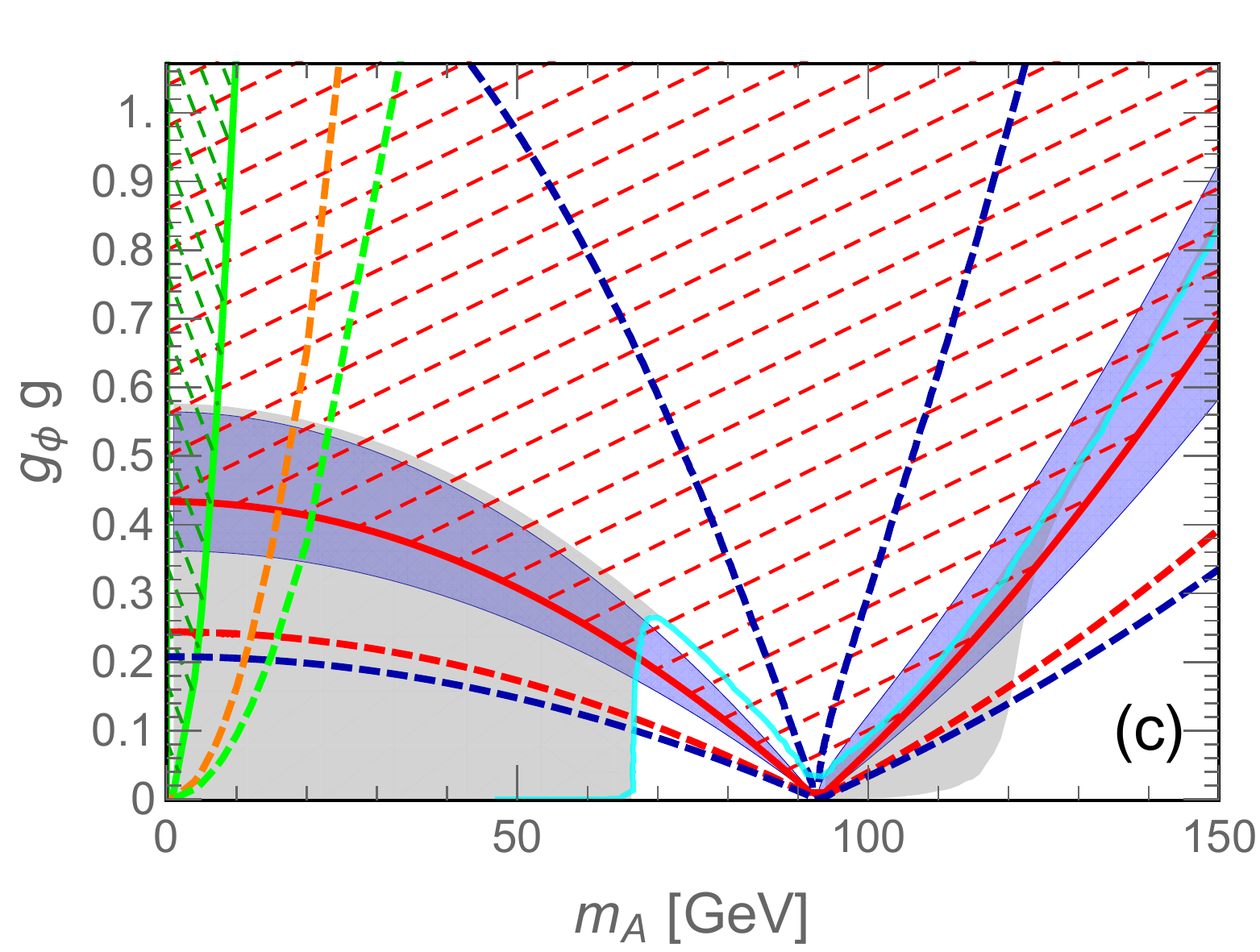}
\caption{Allowed parameter regions in the ($m_A, g_\phi g$) plane, where $m_\phi=46.4$~GeV is used. The blue shaded region is $3\sigma$ C.L. allowed by GC gamma-ray excess data in the scenario {\bf s1}, where the thick dashed blue lines denote the boundaries if astrophysical uncertainties, ${\cal J}_\gamma$ and $\rho_\odot$, are further considered. The magenta, red, and green hatched  regions are excluded at the 95\% C.L. by the monojet search from the CMS data with 12.9 fb$^{-1}$ at 13 TeV, the Fermi-LAT observations of dSphs, and the PICO-60 ${\rm C F}_3{\rm I}$ measurement, respectively.  The dashed lines with magenta, red,  orange, and green colors  are the 95\% C.L. upper limits due to projected sensitivities for HL-LHC  monojet, dSphs, PICO-500L  ${\rm C}_3 {\rm F_8}$, and PICO-500L ${\rm C} {\rm F_3} {\rm I}$. 
In general, the thermal relic density can be accounted for by the gray regions. In (a), (b), and (c), the monojet constraints and thermal relic density shown in the black, brown, and cyan solid curves  correspond to the chosen values: $g=4\pi$, $g=2$,  and $g_\phi=4\pi$, respectively.}
\label{fig:combine-s1} 
\end{center}
\end{figure}

In Fig.~\ref{fig:combine-s1}, taking an illustrative DM mass  of $m_\phi=46.4$~GeV, we show results  in the $(m_A, g_\phi g)$ plane, where the GC gamma-ray excess allowed region at the $3\sigma$ C.L. is given for the scenario {\bf s1}.  There, adopting $g=2$, we show the  (hatched magenta) region excluded by  the very recent CMS  monojet search with 12.9 fb$^{-1}$ of data at 13 TeV, and the projected limit (magenta dashed curve) for the high luminosity LHC (HL-LHC) with an integrated luminosity of 3000 fb$^{-1}$ at 14~TeV \cite{atlas:2016HLLHC}. 
The current CMS monojet constraint is shown to be much less restrictive for this model.
Because we have assumed that the mediator-quark couplings are proportional to the quark's mass, the production of $A$, mainly via the top loop, is dominated  by gluon fusion at the LHC. Therefore, we may expect that the S/B (number of signal to number of background ratio) is approximately the same as the present value, such that the projected limit  on  $g_\phi g$ corresponds to an improvement by a factor $\sim 4.3$.   As shown in Fig.~\ref{fig:combine-s1}, the projected HL-LHC limit could constrain the favored parameter region. However, the projected limit might become less restrictive if $g$ is much different from 2.

\subsection{Relic abundance and combined analyses}\label{sec:relic}

\noindent{\bf Scenario s1}

 In the scenario {\bf s1}, we consider that the DM particles annihilate only via an $s$-channel mediator to the SM quark pair at a velocity $v\sim 10^{-3} c$. 
In Fig.~\ref{fig:combine-s1}, in addition to  the parameter region favored by the GC gamma-ray excess and the 95\% C.L. upper limits placed by LHC monojet searches, as discussed previously, we also show regions excluded from the PICO-60 ${\rm CF}_3{\rm I}$ measurements (hatched green) and  Fermi-LAT dSphs observations (hatched red).
The PICO Collaboration has  proposed a ton-scale PICO-500L detector, having an active volume of about 800~L \cite{PICO500:project,PICO500:projectTalk}. The predicted bound from the projected sensitivity of the PICO-500L, assuming an exposure of 500 kg$\cdot$yrs and the same detection efficiency used in PICO-60, is plotted in the dashed orange or dashed green line in Fig.~\ref{fig:combine-s1},  if the bubble chamber is filled  with ${\rm C}_3 {\rm F_8}$ or ${\rm C} {\rm F_3} {\rm I}$ \cite{Pullia:2014vra}.  The projected PICO experiments will constrained the region with $m_A\lesssim 25$~GeV. However, this region is already excluded by the combined analyses of the relic abundance and the observed dSphs (see Fig.~\ref{fig:combine-s1} and also discussions below).

Note that the DM annihilation cross section obtained in the GC excess could be revised by a factor $\in [0.33, 5.88]$ due to  uncertainties of  the DM profile near the Galactic center and local DM density; including these uncertainties, the new GC excess boundaries are denoted by the thick dashed blue lines in Fig.~\ref{fig:combine-s1}. 

As for the relic density constraint,  $\Omega_{\rm{DM}}  h^2=0.1198 \pm 0.0026$ \cite{PDG,Ade:2013zuv}, for comparison, we show all allowed parameter regions in Fig.~\ref{fig:combine-s1}, considering the parameter range of $g_\phi, g \in [0,4\pi]$. If $m_A> m_\phi$ and the temperature at the decoupling time is not high enough to produce the on-shell $A$, i.e., $m_A> m_\phi +K_\phi$  (with $K_\phi$ the thermally kinetic energy of $\phi$), then the DM annihilation is only relevant to the $s$ channel $\phi \phi \to {\bar q q}$, dominated by the $b {\bar b}$ pair in the final state. We find that only the parameter region with  $2m_\phi < m_A \lesssim 2.7 m_\phi$ is compatible with the combination of all data as well as the GC gamma-ray excess within errors;
near the resonance region, if using the canonical astrophysical parameter $\bar{J}_\Omega^{\mathrm{c}}$ as input, we need to adopt a much larger value of $g_\phi (\gtrsim 10)$ to have a large width of the mediator and  then to suppress the resonant enhancement on the annihilation cross section, so that we can have a good fit to the combination of the relic density and GC excess,  otherwise, we need a large revision to the adopted astrophysical  parameters for accounting for a smaller coupling $g_\phi$. The projected sensitivity of dSphs can strongly constrain this allowed region, as shown in Fig.~\ref{fig:combine-s1}.

For  $m_A < m_\phi +K_\phi$, the $s$ channel $\phi \phi \to {\bar q q}$ and ($t$,$u$) channels $\phi \phi \to AA$ are relevant to the relic abundance (as well as the GC gamma-ray excess), because these two annihilation processes are $s$ wave.   To investigate the channel dependence of the relic abundance, we display results  in Fig.~\ref{fig:combine-s1} for the three coupling values of (i) $g=4 \pi$, (ii) $g=2$, and (iii) $g_\phi=4 \pi$, respectively.  Only the case (i), dominated by $\phi \phi \to {\bar q q}$, is consistent with the GC excess allowed region under the requirement of the scenario {\bf s1}, but, however, is completely excluded by the gamma-ray measurement from dSphs. 

Although cases (ii) and (iii) seem to be reconciled with the observed dSphs and GC gamma-ray excess in the ($m_A, g_\phi g$) plane, they  however  contain a sizable contribution from the DM annihilation into on-shell mediators. This contribution is comparable with that from the DM $s$-channel annihilation into the $b \bar{b}$ pair, in contrast with the assumption of the scenario {\bf s1}.  This may imply that for  $m_A < m_\phi$, the DM annihilation into on-shell mediators plays an important role in the phenomenology of the GC gamma-ray excess.

In concluding this subsection, it is interesting to note that  the constraints arising from $ \bar{t} t + {\not\!\!{E}}_{\rm T}$ (with  $A\to {\not\!\!{E}}_{\rm T}$) channel  and the mediator's visible decay channels at the LHC: $ p p \to A \to \tau^+ \tau^-, \gamma\gamma$,  could be comparable with and/or complementary to the monojet result.\footnote{ The $A\to \tau\tau$ channel is irrelevant to our present case, because we consider the mediator couples only to the quark sectors.} This was recently stressed by Banerjee {\it et al.} \cite{Banerjee:2017wxi}.  

\vskip1.2cm

\noindent{\bf Scenarios s2 and s3}

\begin{figure}[t!]
\begin{center}
\includegraphics[width=0.39\textwidth]{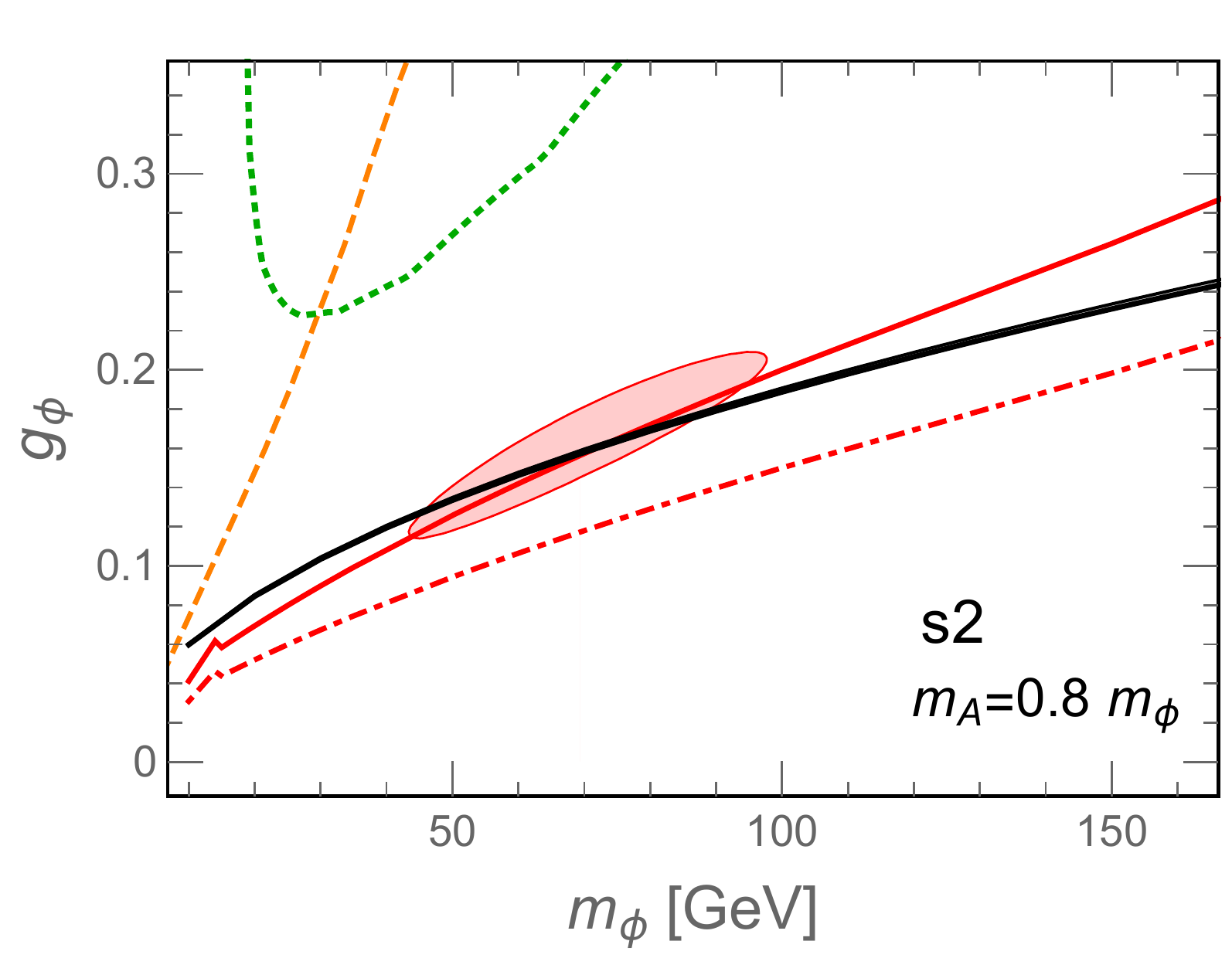}\\
\includegraphics[width=0.39\textwidth]{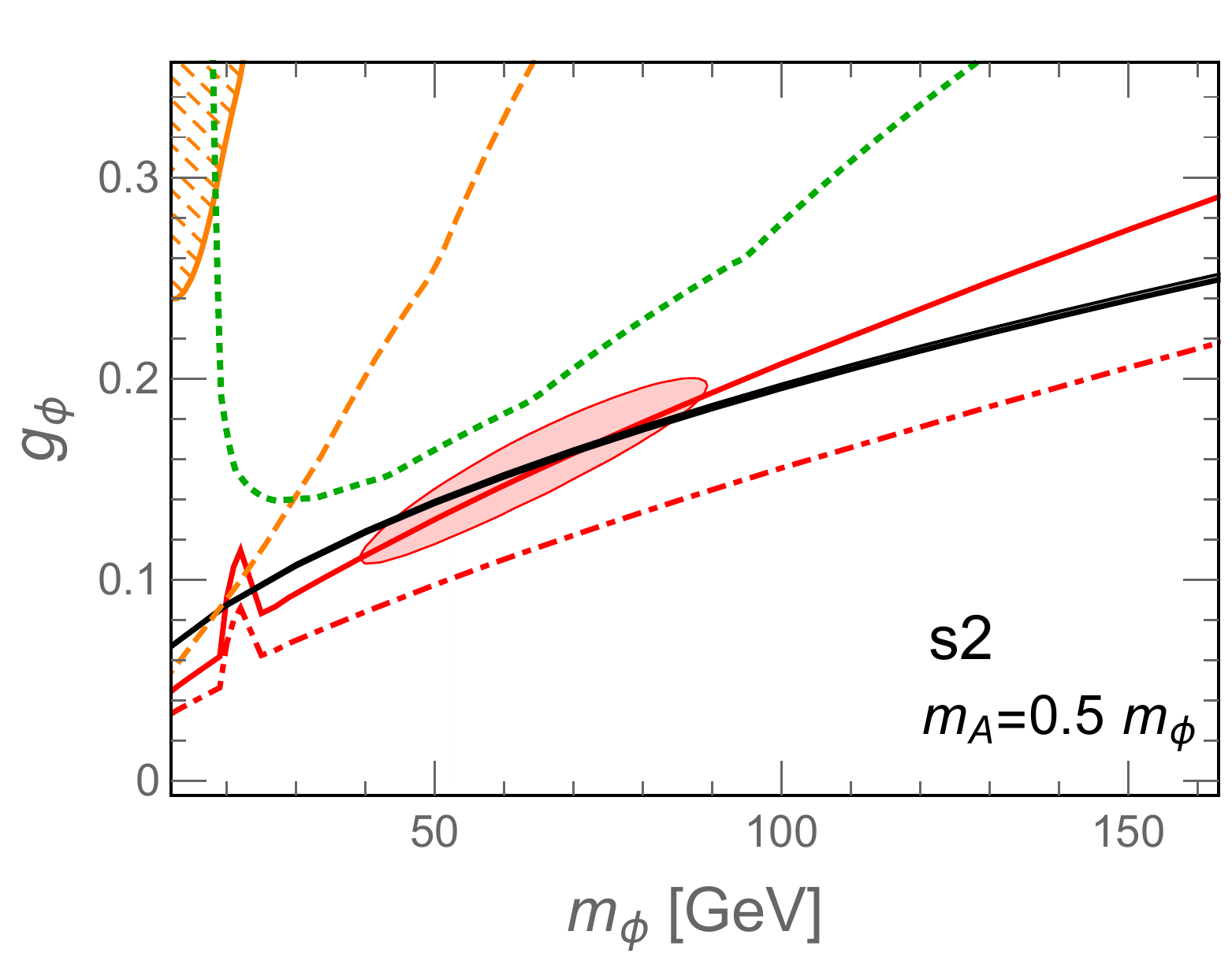} \hskip0.8cm
\includegraphics[width=0.39\textwidth]{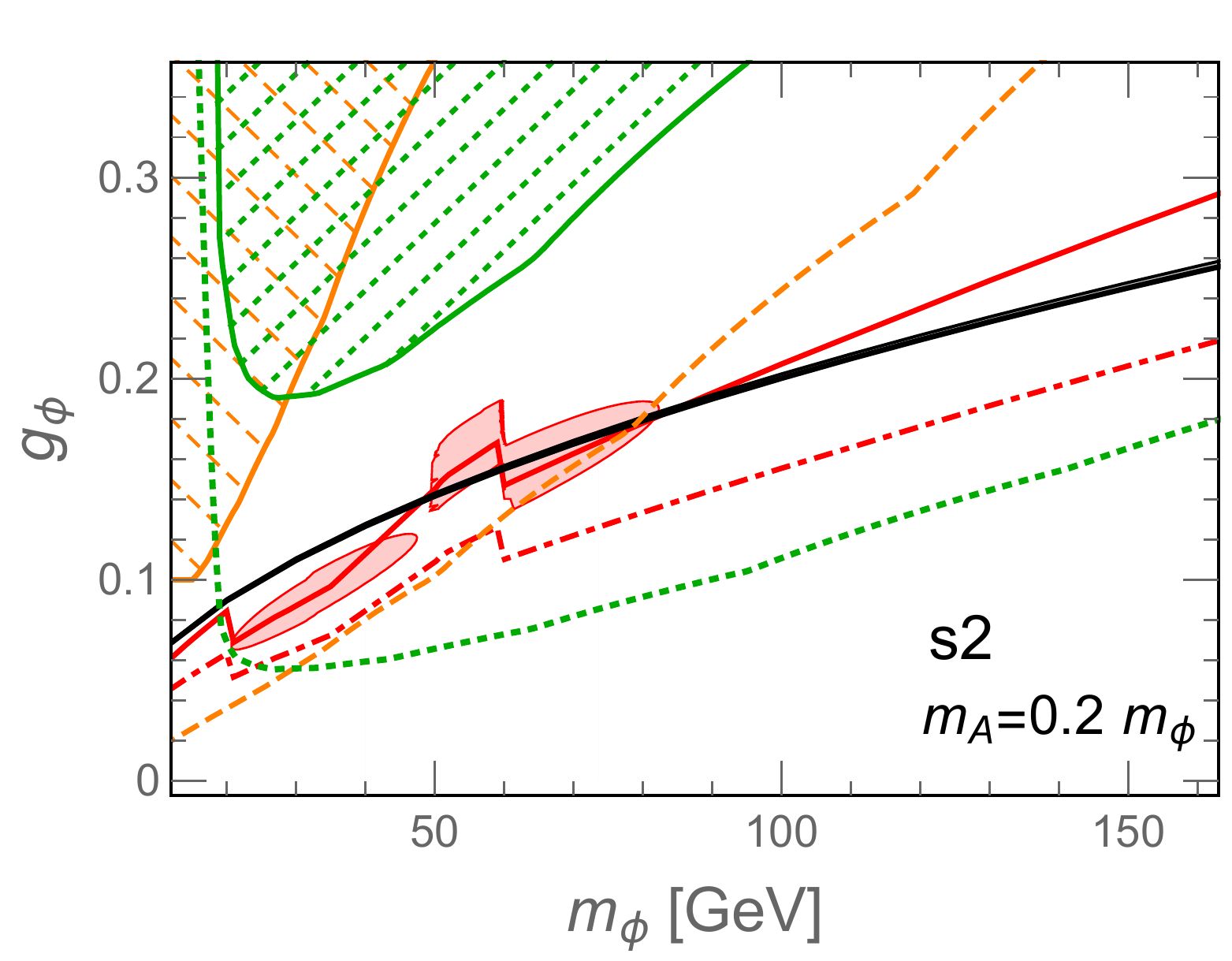}
\caption{Allowed parameter regions in the $(m_\phi, g_\phi)$ plane, for three values of $m_A/m_\phi=0.8, 0.5, 0.2$ in the scenario {\bf s2}. The red shaded region delineated with red line is allowed by $3\sigma$ C.L. fit to  the GC gamma-ray excess data. The red solid curve is the 95\% C.L. upper limit from the Femi-LAT observations of dSphs, while the corresponding dot-dashed curve is the projected 95\% C.L. limit. The orange hatched and green hatched regions are excluded by the PICO-60 ${\rm C}_3 {\rm F_8}$ and PICO-60 ${\rm C} {\rm F_3}I$, respectively, while the orange dashed or green dotted line is the projected 95\% C.L. (upper) limit for PICO-500L with a 500 kg$\cdot$yr exposure if the bubble chamber is filled  with ${\rm C}_3 {\rm F_8}$ or ${\rm C} {\rm F_3} {\rm I}$.  The correct relic abundance is denoted by the black solid curve. }
\label{fig:combine-s2}
\end{center}
\end{figure}

\begin{figure}[t!]
\begin{center}
\includegraphics[width=0.39\textwidth]{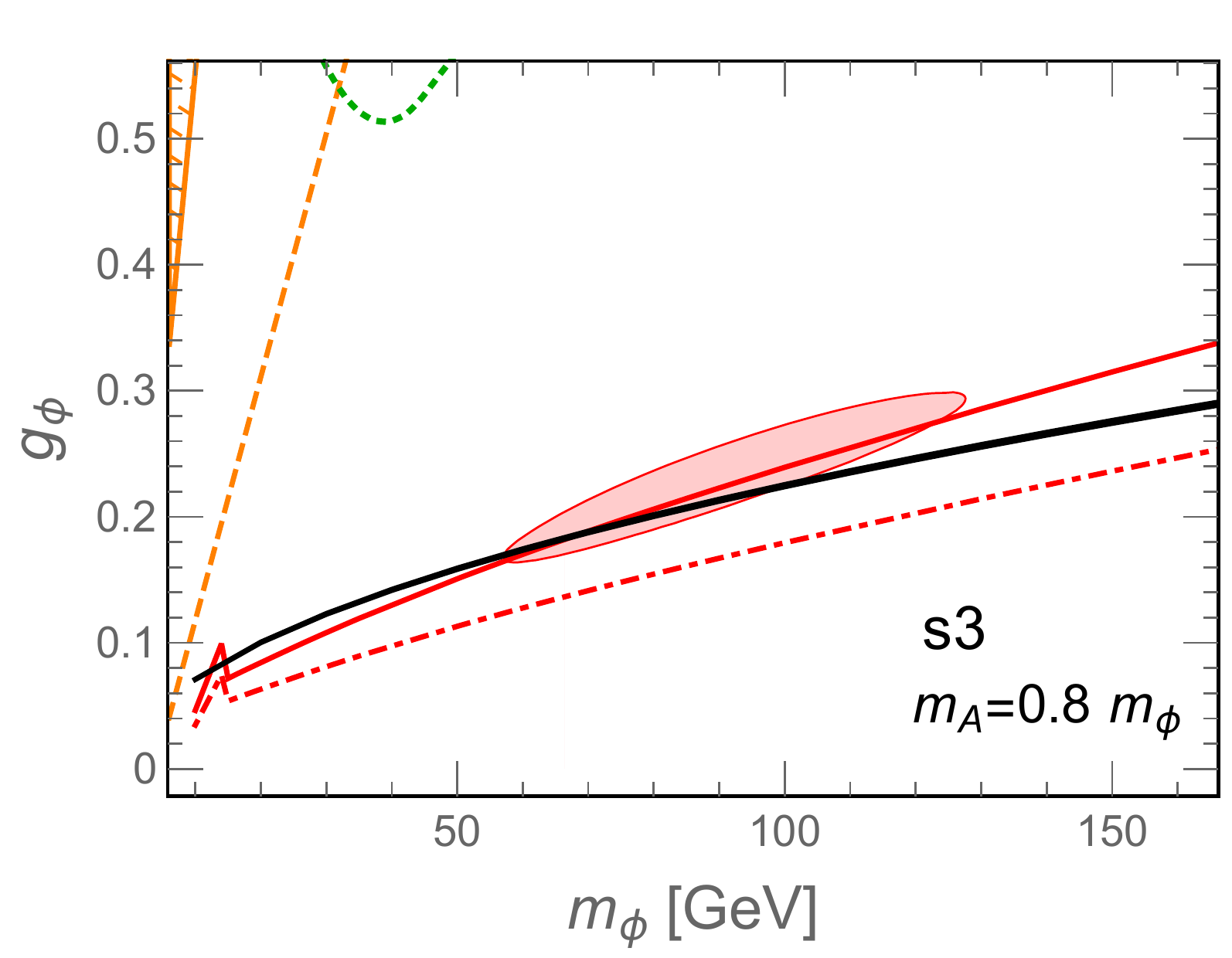}\\
\includegraphics[width=0.39\textwidth]{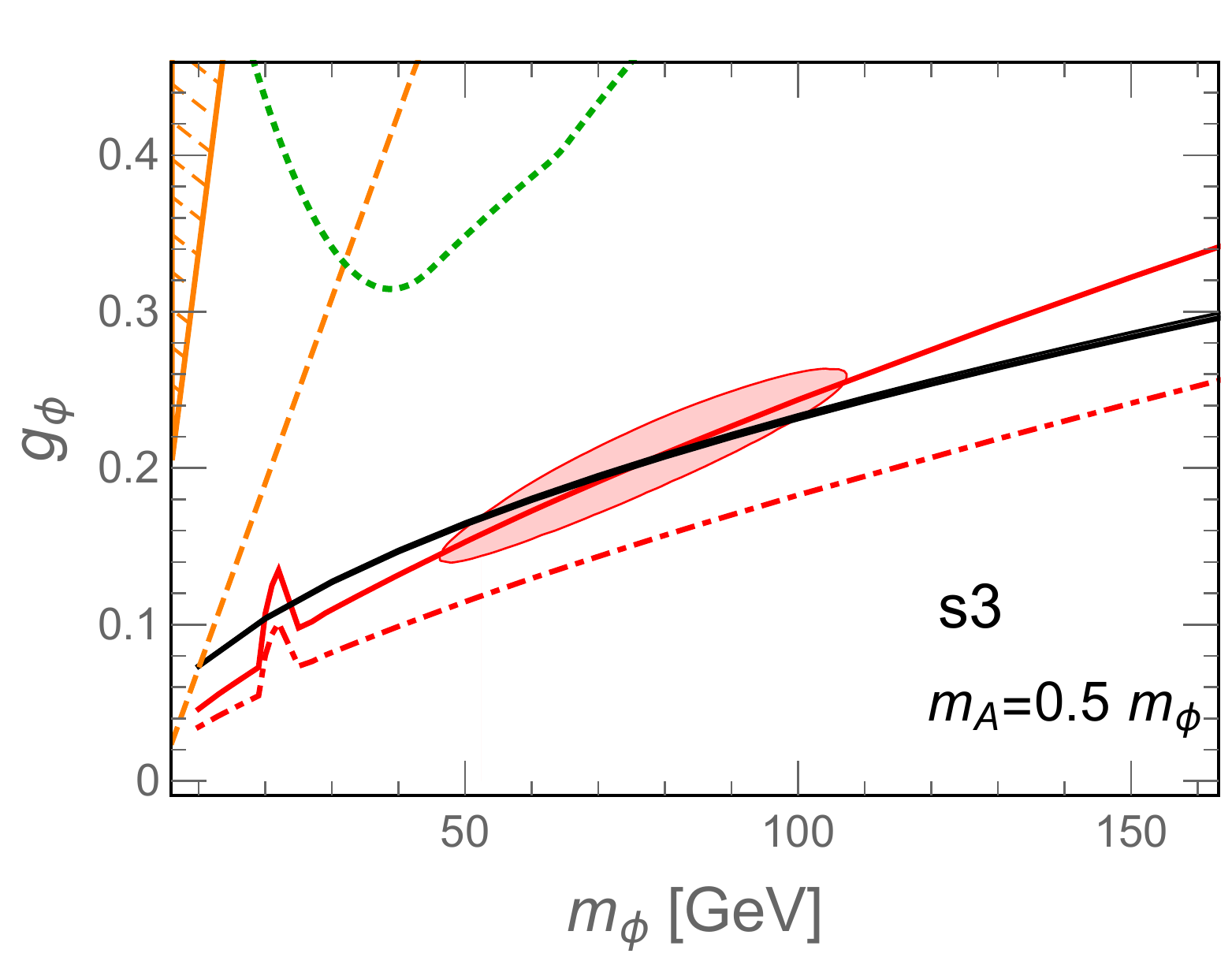} \hskip0.8cm
\includegraphics[width=0.39\textwidth]{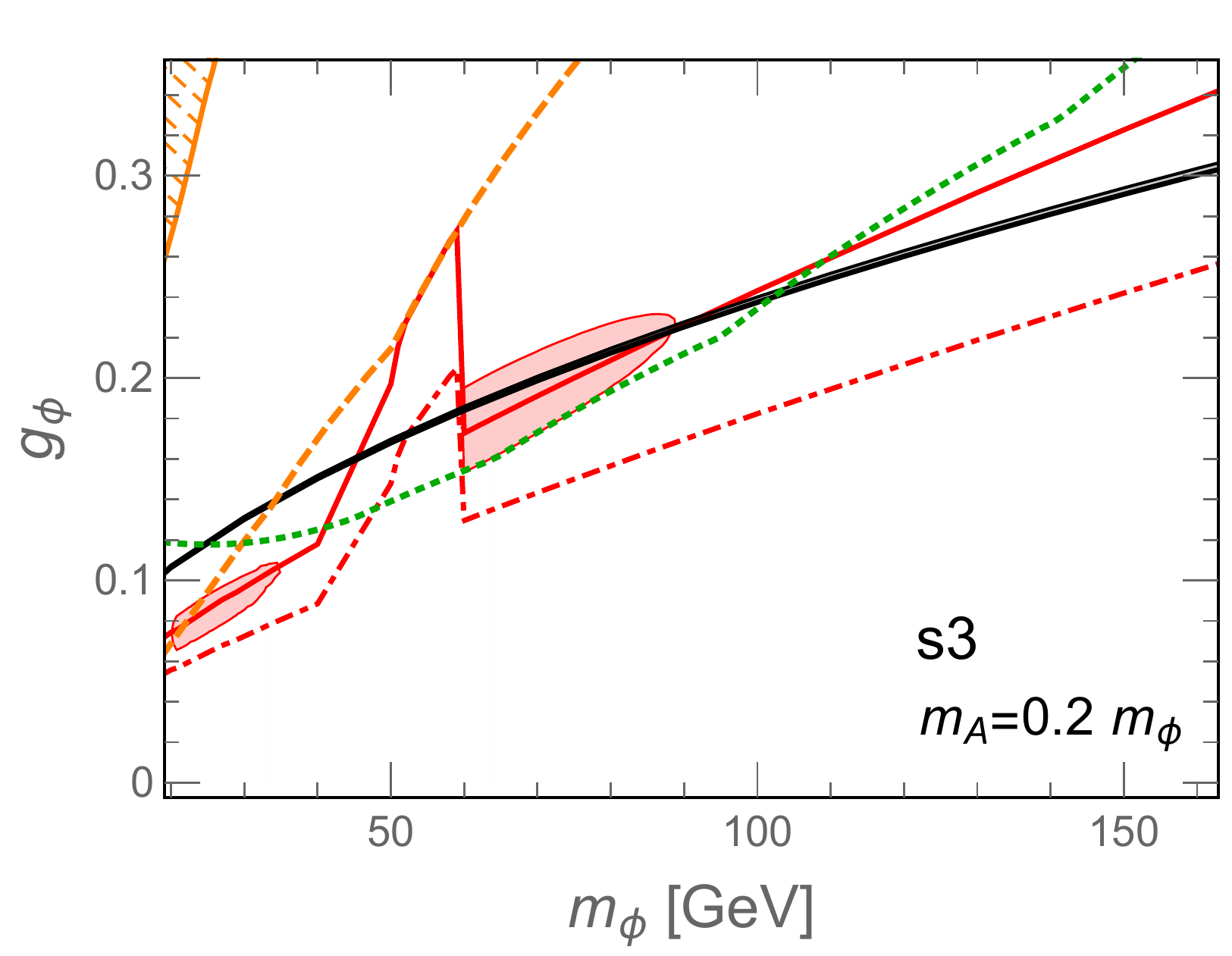}
\caption{Same as Fig.~\ref{fig:combine-s2} but for the scenario {\bf s3}.}
\label{fig:combine-s3}
\end{center}
\end{figure}

In Figs.~\ref{fig:combine-s2} and \ref{fig:combine-s3}, we, respectively, consider two alternative scenarios, {\bf s2} and {\bf s3}, for which, when DM particles move  at an average velocity $v\sim 10^{-3} c$, the former is described by $\langle \sigma v\rangle_{AA} =\sum_q \langle \sigma v \rangle_{\bar q q} $ for the DM annihilation around the GC, and the latter is assumed to respect $\langle \sigma v\rangle_{AA} =20\sum_q \langle \sigma v \rangle_{\bar q q} $.
The constraints due to various experiments are given in the ($m_\phi, g_\phi$) plane, where $g_\phi$ is the only coupling relevant to the DM annihilation into the on-shell mediator pair. We also display results for the three values of $m_A/m_\phi = 0.2, 0.5, 0.8$ to illustrate the mass dependence of the mediator. 

In the scenario {\bf s1}, the value of the low-velocity $\langle \sigma v\rangle_{\bar{b}b}$ is consistent with the thermally averaged cross section required by the relic density ($\sim 1.78\times 10^{-26}\text{cm}^3/s$ corresponding to $m_\phi/T_f\simeq 20$ with $T_f$ the freeze-out temperature for the real scalar DM), whereas, when the $\phi \phi \to AA$ channel is open and $m_A >2m_b$, the fitted cross section becomes larger for the GC gamma-ray excess as shown in Figs.~\ref{fig:GC}, \ref{fig:combine-s2}, and  \ref{fig:combine-s3}. For the latter, at the face value, one may worry the resulting relic density is too small.  However, we find that the thermally averaged annihilation cross sections for $\langle \sigma v\rangle_{AA}$ and  $\sum_q \langle \sigma v \rangle_{\bar q q} $ at the freeze-out temperature are, respectively, smaller by a factor of $\sim 0.77$ and $\sim 0.75$, compared to the corresponding low-velocity annihilation ones. Thus, we can obtain the parameters that produce the correct relic abundance and also provide a good fit to the GC gamma-ray excess. The relevant formulas are collected in Appendices \ref{app:relic} and \ref{app:annXS}. 
     
We observe that the allowed parameter regions, constrained by the current measurements, are, respectively, in the ranges of $g_\phi \in [0.1, 0.2]$ and $ [0.15, 0.25]$ for scenarios {\bf s2} and {\bf s3}, where the constraints on the mediator coupling to quarks are correspondingly in the ranges of  $g \in [2.3, 3.7]$ and $ [0.7, 1.1]$.  
Because the hidden mechanism that suppresses signals for direct detections and colliders is mainly due to the structure of the mediator interacting with the SM quarks via pseudoscalar couplings, the coupling $g$ can be still of order ${\cal O}(1)$, such that the parameter space of interest can be reachable in the PICO-500L measurement.

In Figs.~\ref{fig:combine-s2} and \ref{fig:combine-s3}, we also show the predicted constraints from the projected sensitivities for the gamma-ray observations of dSphs and the PICO-500L experiment. We find that the Fermi-LAT projected observations of dSphs with 15-year data collection (dot-dashed red line) can stringently constrain the most parameter space allowed by the GC gamma-ray excess data, even including the astrophysical uncertainties which are not shown in the plots. In this model, the scattering cross section in the direct detection experiments is suppressed by two powers of momentum transfer. Although, the present PICO-60 results cannot provide sufficient constraints on the parameter space, the PICO group is planning to run PICO-500L using ${\rm C}_3{\rm F}_8$ as the target material, for which ${\rm CF}_3{\rm I}$ can be a substitution \cite{PICO500:project,PICO500:projectTalk,Pullia:2014vra}.  The constraints from PICO-500L, assumed to have a run of 500 kg$\cdot$yr exposure, can be considerably more restrictive for a light mediator with $m_A\lesssim 0.5 m_\phi$ in scenario {\bf s2} and $m_A \lesssim 0.2 m_\phi$ in the scenario {\bf s3}, if, in particular, the detect chamber is filled with ${\rm CF}_3{\rm I}$ (dashed green line).

\section{Conclusions}\label{sec:conclusions}

 In light of observations of  the GC gamma-ray excess, we have investigated a simplified model, in which the scalar dark matter interacts with SM quarks through a pseudoscalar mediator, for which the $s$-wave DM annihilation can occur through an $s$-channel pseudoscalar exchange into the quark pair, or  directly into two on-shell mediators if kinematically allowed. 
 
If the contribution of the low-velocity DM annihilation is mainly due to  the interaction through an $s$-channel off-shell mediator, we have found that only the parameter region with  $2m_\phi < m_A \lesssim 2.7 m_\phi$  is allowed by the combination of all data. For the allowed parameter space near the resonance region,
if using the canonical astrophysical parameter $\bar{J}_\Omega^{\mathrm{c}}$ as input, we need to have  a much larger width of the mediator, corresponding to  $g_\phi \gtrsim 10$, to suppress the so-called resonant enhancement on the annihilation cross section, such that a consistently good fit to the GC excess and relic abundance can become likely; 
otherwise, the canonical astrophysical parameters need be revised largely for accounting for a smaller coupling $g_\phi$.

For the region $2m_b<m_A < m_\phi$ (scenarios {\bf s2} and {\bf s3}), although a larger low-velocity annihilation cross section is obtained to fit the GC gamma-ray excess, the thermally averaged annihilation cross sections at the freeze-out temperature are smaller by a factor of about 0.75$-$0.77 compared to the low-velocity annihilation ones. As a result, we find  that the DM annihilation into two {\it hidden} on-shell mediators, which may be accompanied by an $s$-channel annihilation into the $b {\bar b}$ pair via an off-shell mediator, can be capable of accounting for the GC gamma-ray excess and relic abundance, and evade the current constraints from direct detections, observations of dSphs, and monojets results at the LHC.  In this model, the signal suppression of the hidden sector is mainly due to the coupling of the pseudoscalar mediator to the SM quarks.   

The current constraint from the CMS monojet plus missing transverse energy search are shown to be very weak for this model.
The projected sensitivity of the monojet search at the high luminosity LHC  could constrain the favored region $m_A \gtrsim 2 m_\phi$, where only the DM $s$-channel annihilation into the ${\bar{b} b}$ pair is relevant to the GC gamma-ray excess.

We have shown the regions disfavored by the observation of dSphs, which provide the leading constraints on the GC gamma-ray excess.
Moreover, the projected sensitivity of the 15-year Fermi-LAT observations of dSphs can set a stringent constraint on the most parameter space allowed in this model.

For direct detections, we have presented the exclusion limits of the current LUX WS2014-16 and PICO-60. The latter is especially insensitive to the choice of parameter set, and, moreover,  gives the most stringent exclusion bound among current direct detection experiments.
If the dark matter annihilation is contributed by the on-shell mediator channel over 50\%, this model with a light mediator $m_A\lesssim 0.5 m_\phi$ can be accessible in the projected PICO-500L experiment in the near future.

\acknowledgments \vspace*{-1ex}
 This work was supported in part by the Ministry of Science and Technology of R.O.C. under Grant No.~102-2112-M-033-007-MY3 and  No.~105-2112-M-033-005.

\appendix
\section{Relic abundance}\label{app:relic}

 The Boltzmann equation for $\phi$ with the number density $n_\phi$ is
 \begin{equation}
 a^{-3} \frac{d (n_\phi a^3)}{dt}  =
\langle \sigma v_{\text{M\o l}}\rangle  \left[ (n_\phi^{(0)})^2 - n_\phi^2   \right] \,,
\nonumber   \label{eq:boltzmann}
 \end{equation}
 where  $\langle \sigma v_{\text{M\o l}}\rangle$  is the thermally averaged annihilation cross section, $n_\phi^{(0)}$ is the equilibrium number density of $\phi$, and $v_{\text{M\o l}}$ is the M{\o}ller velocity.  Solving the Boltzmann equation, one can obtain
 the thermal DM relic abundance ($ \Omega_{\rm DM} h^2$) and freeze-out temperature ($T_f=m_\phi /x_f$), given by \cite{Griest:1990kh,Gondolo:1990dk},
\begin{eqnarray}
 \Omega_{\rm DM} h^2 \simeq  \eta\frac{1.04 \times 10^9\  {\rm GeV}^{-1}}{ J \sqrt{g_*} m_{\rm pl} }, \qquad
 x_f\simeq \ln \frac{0.0382 m_{\rm pl} m_{\phi} \langle \sigma v_{\text{M\o l}}\rangle \delta(\delta+2) }{\sqrt{g_*  x_f}},
 \label{eq:xf}
  \end{eqnarray}
where 
\begin{equation}
 J= \int_{x_f}^\infty \frac{\langle \sigma v_{\text{M\o l}}\rangle}{x^2} dx , 
\end{equation}
 $\eta$ = 2 (or 1) for the complex (or real) scalar DM particle,  $m_{\rm pl}\simeq 1.22\times10^{19}$ GeV is the Planck mass, $x_f \equiv m_\phi/T_f \approx 20$, $\delta \equiv n_\phi (x_f)/n_\phi^{(0)}(x_f) - 1$,  and $g_*$ is the number of relativistic degrees of freedom (dof).  A convenient choice is $\delta(\delta+2)=(n+1)$, where $n=0$ corresponds to  the $s$-wave annihilation  \cite{Griest:1990kh} and is relevant to our present model. We use  $g_*\approx 87.25$, which is the sum of the relativistic dof of the $A$ particle and SM for $4~{\rm GeV}<T_f <80~ {\rm GeV}$.   The value for the DM density is $\Omega_{\rm{DM}}  h^2=0.1198 \pm 0.0026$, coming from global fits of cosmological parameters \cite{PDG,Ade:2013zuv}.

\section{Thermally averaged annihilation cross sections}\label{app:annXS}

The thermally averaged annihilation cross section $\langle \sigma v_{\text{M\o l}} \rangle$ is relevant to the determination of the relic density and indirect detection searches. Considering the relic abundance case, the (nonrelativistic) dark matter particles are assumed to be at rest as a whole in the comoving frame. In this case, $\langle \sigma v_{\text{M\o l}} \rangle$  can be obtained equivalently by performing calculations in the laboratory frame, i.e., $\langle \sigma v_{\text{M\o l}} \rangle =\langle \sigma v_{\rm lab}\rangle$, where  $v_{\text{lab}}$ is the DM relative velocity in the rest frame of one of the incoming particles. 

For a temperature $T\lesssim  3 m_\phi$, the thermally averaged annihilation cross section is given by~\cite{Gondolo:1990dk},
\begin{equation}
\langle \sigma v_{\text{M\o l}} \rangle = \frac{1}{8m_\phi^4 T K_2^2 (m_\phi/T)} \int_{4m_\phi^2}^{\infty}
\sigma  \sqrt{s} (s-4m_\phi^2)K_1 (\sqrt{s}/T) ds,
\end{equation}
where $K_{1,2}$ are the modified Bessel functions and $s$ is the center-of-mass energy squared.
For the DM particles satisfying the condition  $x (\equiv m_\phi /T) \gg 1$, the annihilation cross section can be further approximated as
\begin{equation}
\langle \sigma v_{\text{M\o l}} \rangle
\simeq
\frac{2 x^{3/2}}{\sqrt{\pi}}
 \int_{0}^{\infty} \sigma v_{\text{lab}}  \frac{(1+ 2\epsilon) \epsilon^{1/2} }{(1+\epsilon)^{1/4}}
 \left( 1-\frac{15}{4x} + \frac{3}{16x (1+\epsilon)^{1/2}}   \right)
 e^{-\frac{x\epsilon}{(1+\sqrt{1+\epsilon})/2}} d\epsilon
 \,,
 \label{eq:simpliedthermalXS}
 \end{equation}
where $\epsilon=(s-4m_\phi^2)/(4m_\phi^2)\simeq v_{\text{lab}}^2/4$, and the cross sections, if  kinematically allowed, are given by
\begin{eqnarray}\label{eq:simgav-lab}
&& \sum_q (\sigma v_{\rm lab})_{\phi\phi \to \bar{q} q} 
= \sum_q 
  \left(
\frac{g_{\phi}^2 \, g_q^2\, n_c \, m_{\phi }^2 s}{8 \pi  [ (s- m_A^2)^2+m_A^2\Gamma_A^2 ]  (s- 2 m_\phi^2) } 
 \sqrt{1-\frac{4 m_q^2}{s}} \, \theta (s-4m_q^2) 
\right)     \,, \\
&& (\sigma v_{\rm lab})_{\phi\phi \to AA} = \frac{g_\phi^4 m_\phi^2}{16 \pi  \sqrt{s} \sqrt{s-4 m_{\phi }^2}
   \left(s-2 m_{\phi }^2\right) \left(s-2 m_A^2\right)}
\nonumber\\
&& \times \Biggl[ \frac{\sqrt{s-4 m_{\phi }^2} \sqrt{s-4 m_A^2} \left(s-2 m_A^2\right)}{s m_{\phi }^2-4 m_{\phi }^2 m_A^2+m_A^4}
+ 2\ln \left( \frac{ s-2 m_A^2+ \sqrt{s-4 m_{\phi }^2}  \sqrt{s-4 m_A^2}}{ s-2 m_A^2- \sqrt{s-4 m_{\phi }^2}  \sqrt{s-4 m_A^2}} 
 \right)\Biggr]  \theta(s-4m_A^2) 
  \,,   \nonumber\\
\end{eqnarray}
with  $n_c=3$ being the number of the quark's colors. The former cross section is the $s$-channel process, while the latter contains the $u$ and $t$ channels. For the $A$'s decay width, if  kinematically allowed, we will consider the main channels, $\Gamma_A =\sum_q \Gamma(A \rightarrow \bar{q} q)  +  \Gamma(A\to gg) + \Gamma(A\to \phi \phi)$, where  the partial decay widths are explicitly listed in Eqs.~(\ref{eq:partial-width-1}), (\ref{eq:partial-width-2}), and (\ref{eq:partial-width-3}). As for $m_A\sim \Lambda_{\rm QCD}$,  although its decay width depends on the channels of hadronization \cite{Dolan:2014ska}, it was found that the hadron decay widths can be neglected in the numerical calculation of the annihilation cross section \cite{Yang:2016wrl}. 

In analogy to the relic abundance, for the indirect search case, the dark matter particles can be assumed to be at rest as a whole in the Galactic frame, and $x$ in Eq.~(\ref{eq:simpliedthermalXS})
equals to $2/v_p^2$, with $v_p$ the most probable speed of the dark matter distribution. 
Note that both  $\langle \sigma v_{\text{M\o l}} \rangle_{\phi \phi \to \bar{q}q}$ and  $\langle \sigma v_{\text{M\o l}} \rangle_{\phi\phi \to AA}$  are $s$-wave dominant.

\end{document}